\documentclass[a4paper,11pt]{article}
\usepackage{jheppub}
\usepackage{amsmath,amssymb,graphicx,slashed}
\usepackage[utf8x]{inputenc}

\usepackage{lineno}
\nolinenumbers

\usepackage{orcidlink}
\usepackage{adjustbox}

\allowdisplaybreaks[4]

\title{\bf \boldmath Probing TeV-Scale Inverse-Seesaw Leptogenesis and Majorana Dark Matter in $U(1)_{B-L}$ Models at Multi-TeV Muon Colliders}

\author[a]{Xin-Qiang Li\,\orcidlink{0000-0002-3962-3577},}
\author[a]{Himadri Roy\,\orcidlink{0000-0003-3992-2291},}
\author[a]{Tripurari Srivastava\,\orcidlink{0000-0001-6856-9517},}
\author[a,c]{Ya-Dong Yang\,\orcidlink{0000-0002-9327-0557}}
\author[a]{and Xing-Bo Yuan\,\orcidlink{0000-0002-5647-2563}}

\affiliation[a]{Institute of Particle Physics and Key Laboratory of Quark and Lepton Physics~(MOE), Central China Normal University, Wuhan, Hubei 430079, China}
\affiliation[c]{Institute of Particle and Nuclear Physics, Henan Normal University, Xinxiang 453007, China}

\emailAdd{xqli@mail.ccnu.edu.cn}
\emailAdd{himadri027roy@gmail.com}
\emailAdd{tripurari.sri022@gmail.com}
\emailAdd{yangyd@mail.ccnu.edu.cn}
\emailAdd{y@ccnu.edu.cn}

\abstract{We investigate a predictive and testable framework in which dark matter (DM), heavy-neutrino dynamics, and the baryon asymmetry of the Universe originate from correlated interactions within a local $U(1)_{B-L}$ extension of the Standard Model. Unlike conventional $B-L$ constructions based on the type-I seesaw, we employ an inverse-seesaw mechanism realized through a sterile fermion $S_{1}$ and a complex scalar field $\phi$, whose vacuum expectation value simultaneously generates the masses of the heavy neutrinos $N_{1,2}$ and the Majorana DM fermion $\chi$ via Yukawa couplings. The small lepton-number-violating parameter induced by a higher-dimensional operator leads to naturally light active neutrinos together with TeV-scale heavy neutrinos and sizable active-sterile mixing, yielding distinctive collider signatures unavailable in minimal $B-L$ models. The relic abundance of $\chi$ is governed by annihilation channels mediated by the same scalar and gauge interactions, producing a direct and model-specific correlation between successful leptogenesis and the observed DM relic density. A combined parameter-space analysis incorporating neutrino oscillation data, lepton-flavor-violating processes, direct-detection limits, and collider bounds on $N_{1,2}$ and $Z^\prime$ reveals a narrow yet robust region consistent with all these constraints. Representative benchmark points in this region are examined at a future multi-TeV muon collider. Heavy-neutrino production through electroweak processes yields striking signatures in the dilepton plus missing energy ($2\ell + \slashed{E}_T$) and single-lepton plus di-jet plus missing energy ($1\ell + 2j + \slashed{E}_T$) final states. These channels demonstrate that next-generation muon colliders offer a powerful and complementary probe of the inverse-seesaw origin of neutrino masses, the DM relic density, and the TeV-scale leptogenesis within such an extended $U(1)_{B-L}$ framework.}


\begin{document}
\maketitle
\flushbottom

\section{Introduction}

The discovery of the Higgs boson at the Large Hadron Collider (LHC)~\cite{Aad:2012tfa,Chatrchyan:2012ufa} has completed the particle spectrum of the Standard Model (SM) and confirmed the mechanism of electroweak symmetry breaking that endows fermions and gauge bosons with masses~\cite{Higgs:1964pj,Englert:1964et,Guralnik:1964eu,Higgs:1966ev}. Despite the tremendous success, several fundamental questions remain unanswered within the SM framework. A striking example is the origin of tiny but non-zero neutrino masses, strongly supported by neutrino oscillation experiments~\cite{Super-Kamiokande:1998kpq,SNO:2002tuh} and constrained by cosmology to satisfy $\sum m_\nu < 0.12$~eV~\cite{Aghanim:2018eyx}. Since neutrinos are massless in the SM, this observation provides unambiguous evidence for new physics (NP) beyond the SM.

A minimal and widely studied possibility for generating neutrino masses is the type-I seesaw mechanism~\cite{Minkowski:1977sc,Mohapatra:1979ia,Schechter:1980gr,Schechter:1981cv}, in which heavy right-handed neutrinos generate suppressed Majorana masses for active neutrinos. However, in its conventional form, the type-I seesaw mechanism typically requires large Majorana masses or tiny Yukawa couplings, leading to active–sterile mixing angles far below the sensitivity of current colliders~\cite{Mohapatra:2005wg,Drewes:2013gca}. To obtain testable TeV-scale heavy neutrinos while retaining naturally small neutrino masses, extensions of the fermion sector, such as the inverse-seesaw mechanism, are well motivated~\cite{Mohapatra:1986aw,Mohapatra:1986bd,Bernabeu:1987gr,Gavela:2009cd,Parida:2010wq,Garayoa:2006xs,Abada:2014vea,Law:2013gma,Nguyen:2020ehj}, where small lepton-number–violating (LNV) parameters, rather than ultra-heavy mass scales, account for the lightness of active neutrinos~\cite{Asaka:2005pn,Ma:2006fn,Falkowski:2011xh,Falkowski:2017uya,Hugle:2018qbw,Chianese:2019epo,Liu:2020mxj}.

In addition to the non-vanishing neutrino masses, the observed baryon asymmetry of the Universe (BAU)~\cite{Aghanim:2018eyx},
\begin{equation}
\eta_B = \frac{n_B - n_{\bar{B}}}{n_\gamma} \simeq 6.12 \times 10^{-10},
\end{equation}
cannot be generated within the SM, which lacks sufficient CP violation and the necessary out-of-equilibrium dynamics. Leptogenesis~\cite{Fukugita:1986hr,Covi:1996wh,Roulet:1997xa,Pilaftsis:1997jf,Buchmuller:2005eh,Chun:2007vh,Kitabayashi:2007bs,Prieto:2009zz,Suematsu:2011va,AristizabalSierra:2011ab,Hambye:2012fh,Kashiwase:2013uy,Borah:2013bza,Hamada:2015xva,Zhao:2020bzx}, in which heavy Majorana neutrinos produce a lepton asymmetry that is later converted to a baryon asymmetry by sphaleron processes~\cite{Fukugita:1986hr,Davidson:2008bu,Blanchet:2012bk,Manton:1983nd,Klinkhamer:1984di,Kuzmin:1985mm,Khlebnikov:1988sr,Harvey:1990qw}, provides an elegant explanation. Standard thermal leptogenesis~\cite{Davidson:2002qv} typically operates at very high scales, but in scenarios with quasi-degenerate heavy neutrinos, particularly relevant for the inverse-seesaw mechanism, a resonant enhancement~\cite{Flanz:1996fb,Pilaftsis:1997dr,Pilaftsis:1997jf,Pilaftsis:2003gt,Dev:2017wwc,Chakraborty:2021azg,Chakraborty:2022pcc} of CP violation allows successful leptogenesis at the TeV scale~\cite{Flanz:1996fb,Pilaftsis:1997dr,Pilaftsis:1997jf,Pilaftsis:2003gt,Dev:2017wwc,Deppisch:2015qwa,Cai:2017mow,deGouvea:2013zba,Alekhin:2015byh}. Another major shortcoming of the SM is its inability to account for dark matter (DM)~\cite{Bertone:2004pz, Cirelli:2024ssz}, which constitutes about $27\%$ of the energy density of the Universe according to \textit{Planck}~\cite{Planck:2018vyg}. Thus, models linking DM with the origin of neutrino masses are especially appealing, as they address two seemingly unrelated problems within a unified framework~\cite{Asaka:2005pn,Ma:2006fn,Falkowski:2011xh,Falkowski:2017uya,Hugle:2018qbw,Chianese:2019epo,Liu:2020mxj}.

A simple and well-motivated extension of the SM that naturally connects neutrino masses, DM, and baryogenesis is the introduction of a local $U(1)_{B-L}$ gauge symmetry~\cite{Liu:2024esf,Marshak:1979fm,Mohapatra:1980qe}. Spontaneous breaking of this symmetry generates the Majorana masses required for the seesaw mechanism, introduces a massive $Z^\prime$ gauge boson, and can leave behind a residual discrete symmetry (such as a $\mathbb{Z}_2$ symmetry) that stabilizes the DM particle~\cite{Okada:2010wd,Escudero:2016tzx,Okada:2018ktp}. In such a framework, the interactions responsible for generating neutrino masses also play a central role in DM annihilation processes and in producing a baryon asymmetry~\cite{Iso:2010mv,Heeck:2016oda,Dev:2017xry}. The TeV-scale realization of the $U(1)_{B-L}$ model is particularly attractive: both the heavy neutrinos and the $Z^\prime$ boson can lie within the reach of the collider, while the same gauge and Yukawa interactions control DM freeze-out and leptogenesis~\cite{Okada:2010wd,Escudero:2016tzx,Okada:2018ktp,Iso:2010mv,Heeck:2016oda,Dev:2017xry}. 

In this work, we study an extended $U(1)_{B-L}$ inverse-seesaw model that successfully explains the neutrino mass scale, the baryon asymmetry through resonant leptogenesis, and the DM relic abundance. The model contains two right-handed neutrinos, two gauge-singlet fermions, and a complex scalar whose vacuum expectation value (VEV) breaks the $B-L$ symmetry. The resulting heavy neutrinos and the DM particle naturally reside at the TeV scale, establishing a tight connection between cosmological dynamics and collider phenomenology~\cite{Deppisch:2015qwa,Cai:2017mow,deGouvea:2013zba,Alekhin:2015byh,Chakraborty:2022pcc}. We further analyze the collider signatures associated with the production of heavy neutrinos and the $Z^\prime$ boson at future high-energy muon colliders. Such facilities provide a clean environment in which the small but non-negligible active–sterile mixing of the inverse seesaw can still yield observable leptonic and semi-leptonic final states (see, e.g., Refs.~\cite{Deppisch:2015qwa,Cai:2017mow,deGouvea:2013zba,Alekhin:2015byh} for related studies). Using a cut-based analysis for several benchmark points consistent with cosmological constraints, we will examine the discovery prospects at future high-energy muon colliders with $\sqrt{s}=6$ and 10~TeV~\cite{Han:2021udl}.

The paper is organized as follows. Section~\ref{sec:theo} introduces our theoretical setup and the field content of the model. In section~\ref{sec:pheno}, we present the phenomenological constraints on the model, including neutrino mass fitting, lepton-flavor violation, leptogenesis, and the DM relic abundance. Section~\ref{sec:boltzmannevolution} describes the Boltzmann evolution equations that govern the dynamics of heavy neutrinos, DM, and the $B-L$ asymmetry. The numerical analysis of the viable parameter space is presented in section~\ref{sec:numerical}. In section~\ref{sec:collider}, we explore the collider phenomenology at a multi-TeV muon collider, focusing on the most relevant leptonic and semi-leptonic final states. Finally, in section~\ref{sec:summary} we summarize our main results and discuss the broader implications for future experimental tests of such a unified $U(1)_{B-L}$ framework.

\section{Theoretical Framework}
\label{sec:theo}

The model we are considering here is based on the extended gauge symmetry
\begin{equation}
SU(2)_L \otimes U(1)_Y \otimes U(1)_{B-L} \otimes \mathbb{Z}_2,
\end{equation}
and extends the SM by a complex scalar singlet $\phi$ that spontaneously breaks $U(1)_{B-L}$, a set of gauge-singlet fermions $(N,\,S_1)$, and a fermionic DM candidate $\chi$. The $\mathbb{Z}_2$ symmetry is imposed to stabilize the dark sector: $\chi$ is $\mathbb{Z}_2$-odd, while all SM fields and the remaining new fields are $\mathbb{Z}_2$-even. The field content and charge assignments under $(SU(2)_L,\,U(1)_Y,\,U(1)_{B-L},\,\mathbb{Z}_2)$ are listed as
\begin{equation}
\begin{aligned}
& q_L(2,\,+\tfrac{1}{6},\,+\tfrac{1}{3},\,+),
&&  u_R(1,\,+\tfrac{2}{3},\,+\tfrac{1}{3},\,+),
&&  d_R(1,\,-\tfrac{1}{3},\,+\tfrac{1}{3},\,+),\\[4pt]
& L_L(2,\,-\tfrac{1}{2},\,-1,\,+),
&&  e_R(1,\,-1,\,-1,\,+),
&&  N(1,\,0,\,-1,\,+),\\[4pt]
& H(2,\,+\tfrac{1}{2},\,0,\,+),
&&  \phi(1,\,0,\,+2,\,+),\\
& S_1(1,\,0,\,+1,\,+),
&&  \chi(1,\,0,\,-1,\,-),
\end{aligned}
\end{equation}
where the $\mathbb{Z}_2$-odd assignment of $\chi$ forbids all its decays into lighter $\mathbb{Z}_2$-even states, ensuring its stability on cosmological time scales. This construction is inspired by the study in Ref.~\cite{Liu:2024esf}, but with an extended singlet-fermion sector arranged to realize the inverse-seesaw mechanism~\cite{Mohapatra:1986aw,Mohapatra:1986bd,Bernabeu:1987gr,Gavela:2009cd,Parida:2010wq,Garayoa:2006xs,Abada:2014vea,Law:2013gma,Nguyen:2020ehj}.

The gauge-invariant Yukawa sector relevant for neutrino masses and DM mass generation is given by\footnote{Although the operator \(\phi\,\overline{N^c}N\) is allowed by the gauge symmetry, it can be forbidden by an additional symmetry acting on the sterile-fermion sector. For example, one may impose a lepton-number-like global symmetry under which the $\overline{N^c}N$ Majorana term is not invariant, while the Dirac-type term $\overline{N^c}S_1$ and the small $\mu$ term in the $S_1$ sector are retained. Here we effectively assume this symmetry, so that the heavy sterile states form a pseudo-Dirac pair and the neutrino mass matrix retains the inverse-seesaw structure.}
\begin{equation}
\mathcal{L} \supset
-\,y_\nu\,\overline{L}_L \tilde{H} N
-\,M\,\overline{N^{c}}\,S_1
-\,\frac{\lambda_\mu}{\sqrt{2}}\,\phi^\dagger\,\overline{S_1^{c}}\,S_1
-\,\frac{\lambda_\chi}{\sqrt{2}}\,\phi\,\overline{\chi^{c}}\,\chi
\;+\;\text{h.c.},
\label{eq:yukawa}
\end{equation}
where $\tilde{H}\equiv i\sigma_2 H^\ast$ and $\psi^c\equiv C\overline{\psi}^{\,T}$, with $\sigma_2$ and $C$ denoting the second Pauli matrix and the charge-conjugation operator, respectively. The first term generates the Dirac mass $m_D=y_\nu v_H/\sqrt{2}$ after
electroweak symmetry breaking. The second term provides a large lepton-number-conserving singlet mass scale $M$ that links $N$ and $S_1$. The third term induces a small LNV parameter $\mu=\lambda_\mu v_\phi/2$ once $U(1)_{B-L}$ is broken by $\langle\phi\rangle=v_\phi/\sqrt{2}$. Finally, the last term generates the DM mass $m_\chi=\lambda_\chi v_\phi/2$ and, together with the $\mathbb{Z}_2$ symmetry, ensures the stability of $\chi$.

The scalar sector contains the SM Higgs doublet $H$ and the $B-L$-breaking singlet $\phi$, with VEVs given by $\langle H\rangle=v_H/\sqrt{2}$ and $\langle\phi\rangle=v_\phi/\sqrt{2}$ respectively, where $v_H=246~\mathrm{GeV}$. The renormalizable potential involving these
fields reads
\begin{equation}
V(H,\phi)=
-\mu_H^{2}\,H^\dagger H
-\mu_\phi^{2}\,\phi^\dagger\phi
+\lambda_H (H^\dagger H)^2
+\lambda_\phi (\phi^\dagger \phi)^2
+\lambda_{H\phi} (H^\dagger H)(\phi^\dagger \phi).
\label{eq:potential}
\end{equation}
Minimization of $V$ gives
\begin{equation}
\mu_H^2 = \lambda_H v_H^2 + \frac{1}{2}\lambda_{H\phi}v_\phi^2,
\qquad
\mu_\phi^2 = \lambda_\phi v_\phi^2 + \frac{1}{2}\lambda_{H\phi}v_H^2,
\end{equation}
and the physical spectrum contains one CP-odd scalar $\eta$ corresponding to the imaginary component of $\phi$, as well as two CP-even scalars $h$ and $\rho$ obtained from the mixing of the neutral Higgs component and the real part of $\phi$. Here $h$ is identified as the SM Higgs boson, and the mixing angle is denoted by $\theta$.

\section{Phenomenological Constraints}
\label{sec:pheno}

In this section, we examine the main constraints that shape the viable parameter space of the model. These include the requirement of correctly reproducing the observed neutrino mass and mixing pattern, the bounds from LFV decays, the conditions needed for successful leptogenesis, and the DM relic-density limits. Each of these constraints probes a different part of the Yukawa couplings, masses, and mixings in the model; taken together, they determine the region in which the model remains consistent with current data. We discuss these aspects in the following subsections.

\subsection{Neutrino Mass Fitting}

After spontaneous breaking of the electroweak and $U(1)_{B-L}$ symmetries, the Yukawa interactions in Eq.~(\ref{eq:yukawa}) give rise to neutrino mass terms involving the left-handed SM neutrinos $\nu_L$ and the gauge-singlet fermions $N$ and $S_1$. The relevant mass Lagrangian can be written as
\begin{equation}
\mathcal{L}_m^\nu
=
m_D\,\overline{\nu}_L\,N
+ M\,\overline{N^{\,c}}\,S_1
+ \mu\,\overline{S_1^{\,c}}\,S_1
+ \text{h.c.},
\label{eq:mass_terms}
\end{equation}
where the Dirac and singlet mass matrices are given, respectively, by
\begin{equation}
m_D = y_\nu \frac{v_H}{\sqrt{2}},
\qquad
M \equiv M_{NS},
\qquad
\mu = \lambda_\mu \frac{v_\phi}{2}.
\label{eq:mass_defs}
\end{equation}
Here $v_H$ and $v_\phi$ denote the VEVs of the SM Higgs doublet and the $B-L$–breaking scalar $\phi$, respectively. The parameter $\mu$ originates from the gauge-invariant interaction $\phi^\dagger\,\overline{S_1^{\,c}}\,S_1$, and softly violates lepton number after $U(1)_{B-L}$ breaking.

In the flavor basis
\begin{equation}
\psi^{T} = (\nu_L,\; N^{c},\; S_1),
\end{equation}
the complete neutrino mass matrix takes the characteristic inverse-seesaw form
\begin{equation}
\mathcal{M}_\nu =
\begin{pmatrix}
0 & m_D & 0 \\[2mm]
m_D^{T} & 0 & M \\[2mm]
0 & M^{T} & \mu
\end{pmatrix}.
\label{eq:neutrino_matrix}
\end{equation}
In the phenomenologically relevant limit
\begin{equation}
\|\mu\| \ll \|M\|,
\qquad
\|m_D\| \ll \|M\|,
\end{equation}
the effective light-neutrino mass matrix obtained by block diagonalization reads
\begin{equation}
m_\nu
\simeq
m_D\,M^{-1}\,\mu\,(M^{-1})^{T}\,m_D^{T},
\label{eq:effective_mnu}
\end{equation}
which demonstrates that the smallness of the active neutrino masses is controlled by the LNV parameter $\mu$, rather than by an ultra-heavy seesaw scale $M$. As a result, all singlet fermions can naturally reside at the TeV scale, while remaining compatible with the sub-eV neutrino masses required by neutrino oscillation data~\cite{Esteban:2024eli}.

The heavy neutral-fermion spectrum consists of two quasi-degenerate Majorana states per generation, with masses given approximately by
\begin{equation}
m_{\nu_H}^2
\simeq
M^2 + m_D^2,
\label{eq:heavy_states}
\end{equation}
up to a small splitting of $\mathcal{O}(\mu)$. The hierarchy
\begin{equation}
m_\nu \ll m_D \ll M,
\end{equation}
together with the smallness of $\mu$, ensures moderately large active–sterile mixing, while keeping the heavy states nearly degenerate. This quasi-degeneracy plays a crucial role in the resonant enhancement of CP violation in low-scale leptogenesis, and will be exploited in the phenomenological analysis presented in the latter sections.

\subsection{LFV Constraints}

The flavor mixing in the neutrino sector leads to LFV processes at one loop, such as $\mu \to e\gamma$~\cite{TheMEG:2016wtm}, $\tau \to e\gamma$~\cite{Aubert:2009ag}, and $\tau \to \mu\gamma$~\cite{Aubert:2009ag}, as well as the three-body decays
$\mu \to eee$ and $\tau \to \ell\ell\ell$, with $\ell = e, \mu$. The fact that none of these processes has yet been observed so far puts strong bounds on their branching ratios. These bounds restrict the allowed parameter space in the neutrino sector of the model. We compute the contributions to the LFV processes and use the experimental limits to constrain the model.

\subsection{Leptogenesis}

In our framework, the decay of heavy Majorana neutrinos $N_i$ plays a crucial role in the generation of the BAU~\cite{Plumacher:1996kc,Davidson:2008bu,Fong:2021tqj}. The Yukawa interactions characterized by the coupling $y_\nu$ in Eq.~(\ref{eq:yukawa}) permit the out-of-equilibrium decays 
\begin{equation}
N_i \rightarrow L_\alpha H, 
\qquad 
N_i \rightarrow \bar{L}_\alpha H^*,
\end{equation}
where $L_\alpha$ represents the lepton doublet and $H$ is the SM Higgs field. When the Yukawa coupling $y_\nu$ contains complex phases, these processes can violate CP symmetry, resulting in a small difference between the decay rates of a neutrino and its CP-conjugate channel~\cite{Buchmuller:2004nz}. The CP asymmetry parameter associated with the decays of $N_i$ is defined by
\begin{equation}
\epsilon_i = 
\frac{
\sum_j \Gamma(N_i \to L_\alpha H) - \Gamma(N_i \to \bar{L}_\alpha H^*)
}{
\sum_j \Gamma(N_i \to L_\alpha H) + \Gamma(N_i \to \bar{L}_\alpha H^*)
},
\label{eq:epsilon_def}
\end{equation}
which follows the standard treatment in thermal leptogenesis~\cite{Buchmuller:2002rq,Giudice:2003jh,Buchmuller:2004nz,Chakraborty:2019zas,Rahat:2020mio}.

\begin{figure}[t]
    \centering
    \includegraphics[width=0.40\linewidth]{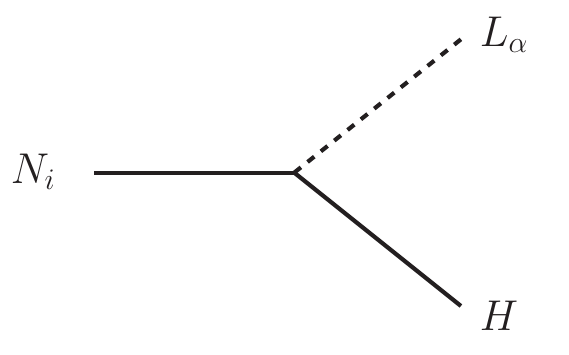}\\[4pt]
    \includegraphics[width=0.40\linewidth]{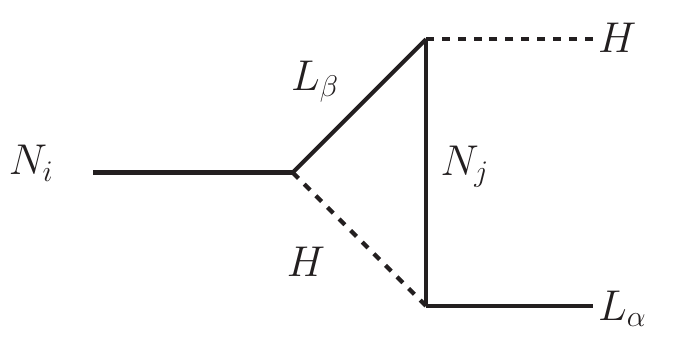}\quad
    \includegraphics[width=0.40\linewidth]{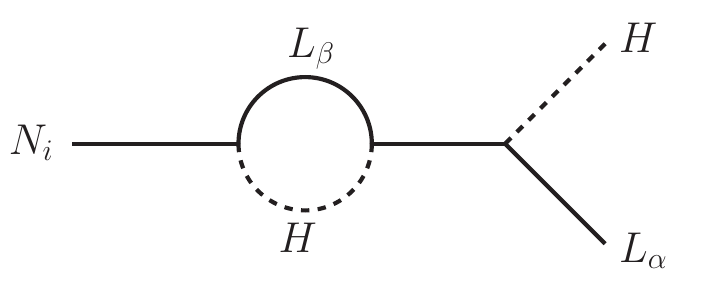}
    \caption{
    Feynman diagrams contributing to the CP-violating decays of heavy Majorana neutrinos. The top diagram shows the tree-level process $N_i \to L_\alpha H$, while the bottom diagrams represent the one-loop vertex and self-energy corrections mediated by virtual heavy neutrino states $N_j$, whose interference with the tree-level amplitude generates the CP asymmetry $\epsilon_i$. \label{fig:leptogenesis_feyn}}
\end{figure}

The decay amplitude of $N_i \to L_\alpha H$ receives contributions from both the tree-level diagram as well as the one-loop vertex and self-energy corrections, as depicted in Fig.~\ref{fig:leptogenesis_feyn}. The interference between these diagrams generates a nonvanishing $\epsilon_i$~\cite{Davidson:2008bu}. In the type-I seesaw picture, the total CP asymmetry arising from these interferences is given by
\begin{equation}
\begin{aligned}
\epsilon_i = - \sum_{j \neq i}
\frac{M_{N_i} \, \Gamma_{N_j}}{M_{N_j}^2} \,
\frac{
\mathrm{Im}\!\left[(y_\nu y_\nu^\dagger)_{ij}^2\right]
}{
(y_\nu y_\nu^\dagger)_{ii} \, (y_\nu y_\nu^\dagger)_{jj}
}
\Bigg\{
&x_{ij}
\left[
(1 + x_{ij})
\ln\!\left(1 + \frac{1}{x_{ij}}\right)
- 1
\right]
\\
& +
\frac{
x_{ij} (x_{ij} - 1)
}{
(x_{ij} - 1)^2 + (\Gamma_{N_j}/M_{N_i})^2
}
\Bigg\},
\end{aligned}
\label{eq:epsilon_expression}
\end{equation}
where $x_{ij} = M_{N_j}^2/M_{N_i}^2$, and $\Gamma_{N_j}$ denotes the total decay width of the heavy neutrino $N_j$~\cite{Pilaftsis:2003gt,Garny:2011hg,Iso:2013lba,Iso:2014afa}. The first term within the brackets originates from the vertex correction, while the second term represents the self-energy (wave-function) contribution.

In the case where the heavy-neutrino mass spectrum is nearly degenerate,
\begin{equation}
\left| M_{N_j}^2 - M_{N_i}^2 \right| \simeq M_{N_i} \Gamma_{N_j},
\label{eq:degenerate_condition}
\end{equation}
the self-energy term in Eq.~\eqref{eq:epsilon_expression} experiences a resonant enhancement~\cite{Hambye:2001eu,Hambye:2004jf,Dev:2017wwc}. Consequently, the CP asymmetry $\epsilon_i$ can become of order unity, even when the Yukawa couplings are relatively small. This phenomenon, known as \textit{resonant leptogenesis}~\cite{Pilaftsis:1997dr,Pilaftsis:1998pd,Branco:2006hz,Brdar:2019iem}, enables the generation of the observed BAU at the TeV scale~--~much lower than the scale required by traditional high-scale thermal leptogenesis.

The asymmetry produced in the decays of $N_i$ is subsequently converted into a net baryon asymmetry through $B+L$-violating sphaleron transitions that remain active above the electroweak phase transition temperature~\cite{Klinkhamer:1984di,Kuzmin:1985mm,Khlebnikov:1988sr}. In this scenario, the interplay between the small mass splitting among the heavy neutrinos and the complex structure of $y_\nu$ determines the magnitude of the CP asymmetry, ensuring successful low-scale leptogenesis consistent with the current experimental data~\cite{Burnier:2005hp,Plumacher:1996kc}.

\subsection{Dark Matter}

The fermionic field $\chi$ serves as the DM candidate in our framework. Its stability is ensured by an imposed discrete $\mathbb{Z}_2$ symmetry, under which $\chi$ is odd, while all SM particles remain even. This discrete charge assignment forbids any renormalizable interaction that could induce its decay into SM states, thereby guaranteeing the cosmological longevity of $\chi$.

To investigate the DM phenomenology, we evaluate the relic abundance of $\chi$ and confront it with the observed DM relic density determined by the \textit{Planck} collaboration~\cite{Planck:2018vyg,Bertone:2004pz,Cirelli:2024ssz}. The numerical analysis is performed using the \texttt{micrOMEGAs} package~\cite{Belanger:2018mqt}, which computes the thermal relic density by solving the Boltzmann equation, including all relevant annihilation and coannihilation channels. The results are further constrained using the latest direct-detection limits reported by the LUX-ZEPLIN (LZ-2024) experiment~\cite{LZ:2024zvo}, ensuring that the parameter space remains consistent with the existing observational data.

\begin{figure}[t]
    \centering
    \includegraphics[width=0.39\linewidth]{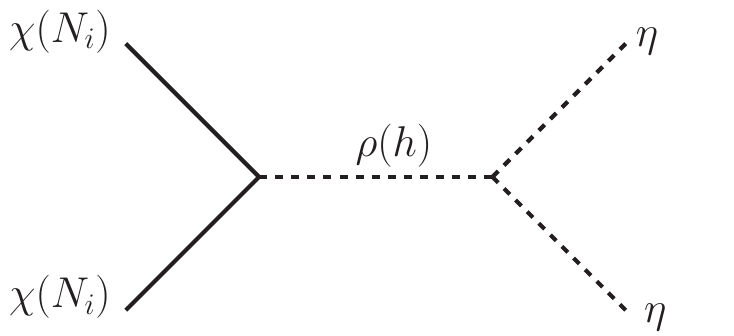}\qquad
    \includegraphics[width=0.30\linewidth]{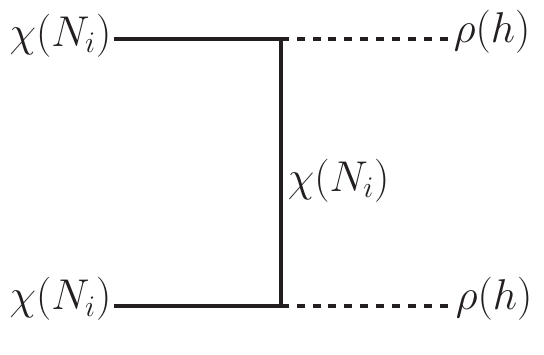}\\[0.5cm]
    \includegraphics[width=0.35\linewidth]{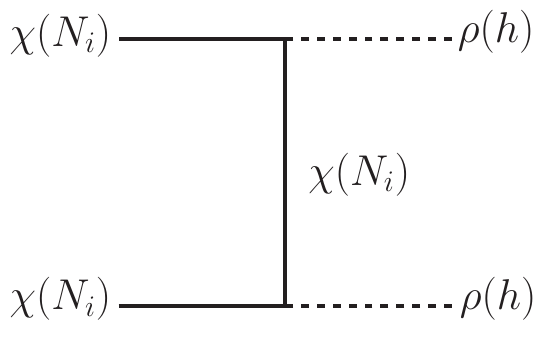}\qquad
    \includegraphics[width=0.35\linewidth]{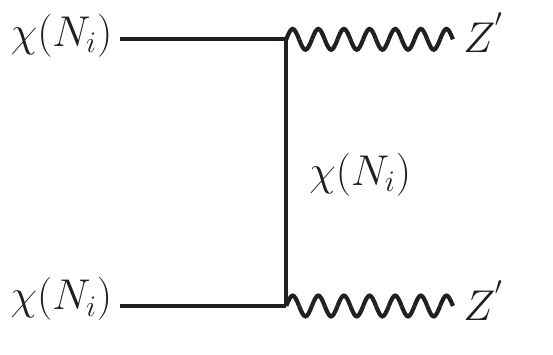}\\[0.5cm]
    \includegraphics[width=0.35\linewidth]{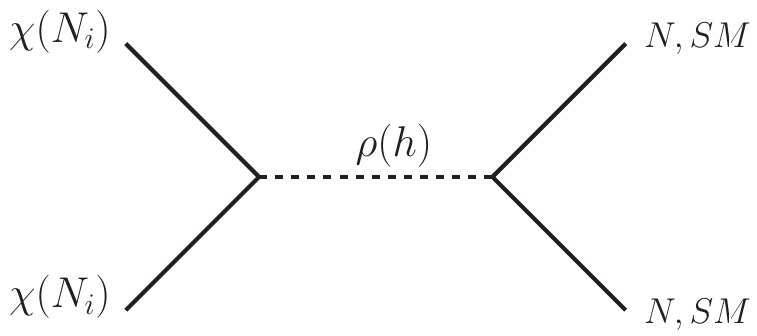}\qquad
    \includegraphics[width=0.38\linewidth]{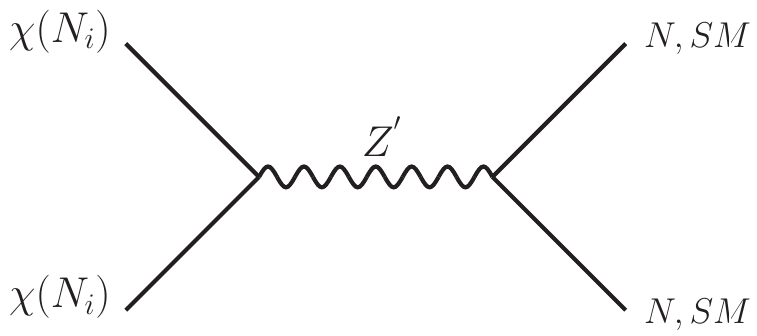}\\[0.5cm]
    \includegraphics[width=0.38\linewidth]{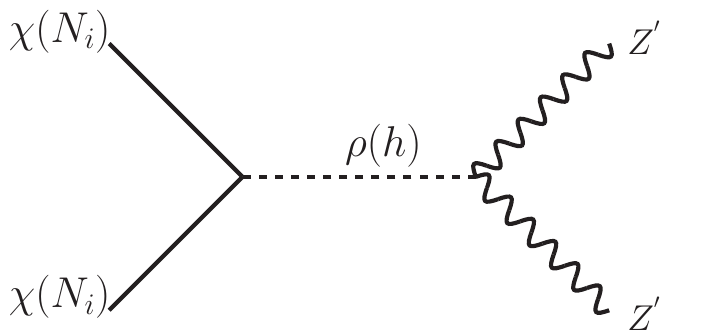}\qquad
    \includegraphics[width=0.38\linewidth]{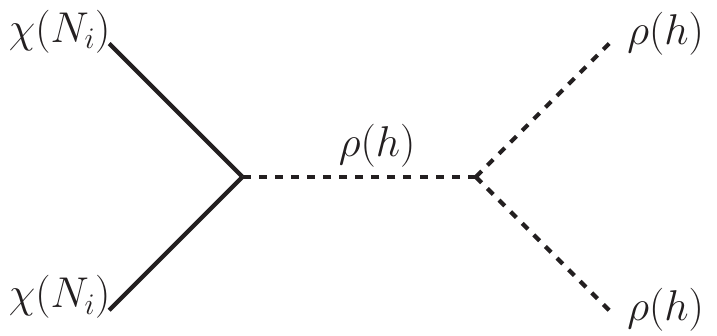}
    \caption{
    Representative annihilation diagrams of the DM particle $\chi$. These include interactions mediated by the scalar fields $\eta$, $\phi$, and by the $U(1)_{B-L}$ gauge boson $Z^\prime$, as well as the mixed annihilation channels with heavy neutrinos. The relic density and freeze-out behavior of $\chi$ are governed by these dominant processes. \label{fig:dmprocessfey}}
\end{figure}

The main annihilation processes that contribute to the thermal freeze-out of $\chi$ are illustrated in Fig.~\ref{fig:dmprocessfey}. The most significant channels can be summarized as follows~\cite{Liu:2024esf}:
\begin{itemize}
    \item Annihilation of $\chi\chi$ into a pair of singlet scalars $\eta\eta$ through the $s$-channel exchange of the scalar mediator $\rho(h)$.

    \item Annihilation of $\chi\chi$ into two scalar mediators $\rho(h)\rho(h)$ through the $t$-channel exchange of $\chi$.

    \item Annihilation of $\chi\chi$ or $N_iN_i$ into $\rho(h)\rho(h)$ through the $t$-channel exchange of $\chi$.

    \item Annihilation of $\chi\chi$ or $N_iN_i$ into $Z'Z'$ through the $t$-channel exchange of $\chi$.

    \item Annihilation of $\chi\chi$ into SM particles or sterile neutrinos $N$ through the $s$-channel exchange of $\rho(h)$.

    \item Annihilation of $\chi\chi$ into SM particles or sterile neutrinos $N$ through the $s$-channel exchange of the gauge boson $Z'$.

    \item Annihilation of $\chi\chi$ or $N_iN_i$ into $Z'Z'$ through the $s$-channel exchange of $\rho (h)$.

    \item Annihilation of $\chi\chi$ or $N_iN_i$ into $\rho(h)\rho(h)$ through the $s$-channel exchange of $\rho(h)$.
\end{itemize}

In this analysis, the mass of $\chi$ is taken to lie above the electroweak scale to avoid the experimental limits on the invisible Higgs decay width. This mass range also allows $\chi$ to achieve the correct relic density through thermal freeze-out, governed primarily by the interplay between the scalar and gauge portal interactions described above.

The viable parameter space consistent with the observed DM relic abundance and direct-detection bounds typically corresponds to moderate Yukawa couplings and a $Z^\prime$ mass of order a few TeV. These conditions ensure that the model remains perturbative and compatible with current collider and cosmological constraints, while simultaneously providing a compelling unified framework for neutrino mass generation and DM stability~\cite{Iso:2010mv,Dev:2017xry}.

\section{Boltzmann Evolution}
\label{sec:boltzmannevolution}

The cosmological evolution of the heavy Majorana neutrinos $N_i$, the DM fermion $\chi$, and the generated $B-L$ asymmetry is described by a coupled system of Boltzmann equations~\cite{Plumacher:1996kc,Davidson:2008bu,Iso:2010mv,Dev:2017xry,Borah:2021mri}. These equations govern the evolution of the comoving number densities $Y_X \equiv n_X / s$, where $n_X$ is the number density of a given particle species and $s$ denotes the entropy density of the Universe. The temperature dependence is expressed in terms of the dimensionless variable $z = m_{N_1}/T$, with $T$ being the plasma temperature. The expansion of the Universe is characterized by the Hubble parameter $H$, evaluated at the relevant temperature scale.

The evolution equations for $Y_{N_i}$, $Y_\chi$, and the net $B-L$ asymmetry $Y_{B-L}$ are given, respectively, by~\cite{Liu:2024esf}
\begin{align}
\frac{dY_{N_i}}{dz} &= 
- \frac{z}{s\,H(m_{N_1})}
\left(\frac{Y_{N_i}}{Y_{N_i}^{\text{eq}}} - 1\right)
\Big(\gamma_{N_i \to \ell H} + 2\,\gamma_{N_i \ell \to q t} + 4\,\gamma_{N_i t \to q \ell}\Big)
\nonumber\\[3pt]
&\quad
- \frac{z}{s\,H(m_{N_1})}
\left[
\left(\frac{Y_{N_i}}{Y_{N_i}^{\text{eq}}}\right)^2 - 1
\right]
\Big(
2\,\gamma_{N_i N_i \to \rho\rho,\,Z^\prime Z^\prime} +
2\,\gamma_{N_i N_i \to \rho Z^\prime,\,hZ^\prime} +
2\,\gamma_{N_i N_i \to \text{SM}}
\Big)
\nonumber\\[3pt]
&\quad
+ \frac{z}{s\,H(m_{N_1})}
\left[
\left(\frac{Y_\chi}{Y_\chi^{\text{eq}}}\right)^2 -
\left(\frac{Y_{N_i}}{Y_{N_i}^{\text{eq}}}\right)^2
\right]
2\,\gamma_{\chi\chi \to N_i N_i},
\\
\frac{dY_\chi}{dz} &=
- \sum_{i=1,2}
\frac{z}{s\,H(m_{N_1})}
\left[
\left(\frac{Y_\chi}{Y_\chi^{\text{eq}}}\right)^2 -
\left(\frac{Y_{N_i}}{Y_{N_i}^{\text{eq}}}\right)^2
\right]
2\,\gamma_{\chi\chi \to N_i N_i}
\nonumber\\[3pt]
&\quad
- \frac{z}{s\,H(m_{N_1})}
\left[
\left(\frac{Y_\chi}{Y_\chi^{\text{eq}}}\right)^2 - 1
\right]
\Big(
2\,\gamma_{\chi\chi \to \rho\rho,\,Z^\prime Z^\prime} +
2\,\gamma_{\chi\chi \to \rho Z^\prime,\,hZ^\prime} +
2\,\gamma_{\chi\chi \to \text{SM}}
\Big),
\\
\frac{dY_{B-L}}{dz} &=
\sum_{i=1,2}
\frac{z}{s\,H(m_{N_1})}
\left[
\varepsilon_i
\left(\frac{Y_{N_i}}{Y_{N_i}^{\text{eq}}} - 1\right)
- \frac{Y_{B-L}}{2\,Y_\ell^{\text{eq}}}
\right]
\gamma_{N_i \to \ell H}
\nonumber\\[3pt]
&\quad
- \sum_{i=1,2}
\frac{z}{s\,H(m_{N_1})}
\frac{Y_{B-L}}{Y_\ell^{\text{eq}}}
\left(
\frac{Y_{N_i}}{Y_{N_i}^{\text{eq}}}\,
\gamma_{N_i \ell \to q t}
+ 2\,\gamma_{N_i t \to q \ell}
\right).
\end{align}
The physical quantities appearing in the above system are defined as follows: 
\begin{itemize}
    \item $Y_X \equiv n_X/s$: comoving number density of the species $X$;
    \item $Y_X^{\text{eq}}$: equilibrium comoving number density;
    \item $H(m_{N_1})$: Hubble expansion rate evaluated at $T = m_{N_1}$;
    \item $\varepsilon_i$: CP asymmetry generated in the decay of $N_i$~\cite{Davidson:2008bu,Pilaftsis:2003gt,Garny:2011hg,Iso:2013lba,Iso:2014afa};
    \item $\gamma_{a \to b}$, $\gamma_{ab \to cd}$: thermally averaged reaction densities for decay and scattering processes~\cite{Plumacher:1996kc,Davidson:2008bu}.
\end{itemize}
Explicit expressions for the thermally averaged $\gamma$ functions can be found in Refs.~\cite{Plumacher:1996kc,Buchmuller:2004nz}.

At early times, all species are assumed to be in thermal equilibrium with the plasma, such that $Y_X = Y_X^{\text{eq}}$ for each particle species $X$. As the Universe expands and cools, the interaction rates of $N_i$ and $\chi$ eventually fall below the Hubble expansion rate, making these species decouple from the thermal bath. The freeze-out of $\chi$ determines its present relic density~\cite{Belanger:2018mqt,Planck:2018vyg}, while the departure of $N_i$ from equilibrium combined with CP-violating decays produces a net lepton asymmetry. Subsequently, this asymmetry is converted into the observed BAU through electroweak sphaleron transitions~\cite{Klinkhamer:1984di,Kuzmin:1985mm,Khlebnikov:1988sr,Burnier:2005hp}, thereby establishing a unified thermal origin for both DM and baryogenesis within the model~\cite{Iso:2010mv,Dev:2017xry,Borah:2021mri}.

\section{Numerical Analysis}
\label{sec:numerical}

To explore the phenomenological viability of our framework, we perform a comprehensive numerical study that covers both the DM and leptogenesis sectors. The model is characterized by the following set of independent parameters:
\begin{equation}
\left\{ m_\rho,\, m_{Z^\prime},\, M_{N_1},\, M_{N_2},\, m_\chi,\, m_{\nu_0},\, v_\phi,\, \theta,\, y_\nu \right\}.
\end{equation}
These parameters denote the masses of the additional scalar $\rho$, the $U(1)_{B-L}$ gauge boson $Z^\prime$, the two heavy Majorana neutrinos $N_{1,2}$, and the DM fermion $\chi$, as well as the light neutrino mass scale $m_{\nu_0}$, the VEV of the scalar field $\phi$, the scalar mixing angle $\theta$, and the Yukawa coupling matrix $y_\nu$, respectively.

We perform a detailed parameter scan to identify the parameter regions consistent with current cosmological and experimental observations. The analysis focuses on reproducing the measured DM relic density, generating the correct baryon asymmetry through low-scale leptogenesis, and satisfying the latest direct-detection bounds from the LZ-2024 experiment~\cite{LZ:2024zvo}. The cosmological observables used as input constraints are
\begin{equation}
Y_B^{\text{obs}} \simeq 8.7 \times 10^{-11},
\qquad
\Omega_{\text{DM}} h^2 \simeq 0.12,
\end{equation}
which correspond to the observed baryon-to-entropy ratio and the DM relic abundance, respectively.

For each point in the scan, we also impose requirements from neutrino oscillation data~\cite{Esteban:2024eli} and LFV processes~\cite{TheMEG:2016wtm,Aubert:2009ag}. The light-neutrino mass matrix is constructed from the chosen heavy-neutrino spectrum and Yukawa couplings, and only those points that reproduce the observed mass-squared differences and mixing angles are kept. For the same parameter set, we solve the Boltzmann equations discussed in the previous section and compute the resulting baryon asymmetry yield $Y_B$. The loop-induced amplitudes for the radiative LFV transitions $\ell_i \rightarrow \ell_j \gamma$ are evaluated to ensure compatibility with the corresponding experimental upper limits. Only points satisfying \emph{all} of these conditions are regarded fully viable.

To remain compatible with collider searches for heavy dilepton resonances~\cite{ATLAS:2019erb, CMS:2021ctt}, we fix the $Z^\prime$ mass to
$m_{Z^\prime} = 7~\text{TeV}$, which lies safely above the current experimental lower limit of $m_{Z^\prime} \gtrsim 5.15~\text{TeV}$. Furthermore, to respect the spin-independent scattering bounds from LZ-2024~\cite{LZ:2024zvo}, the scalar mixing angle is chosen to be small,
$\theta \sim \mathcal{O}(0.01)$, ensuring that the Higgs-portal contribution to direct detection remains well below the current experimental sensitivity while retaining efficient annihilation through the gauge interaction.

\begin{figure}[t]
\centering
\includegraphics[width=0.55\textwidth]{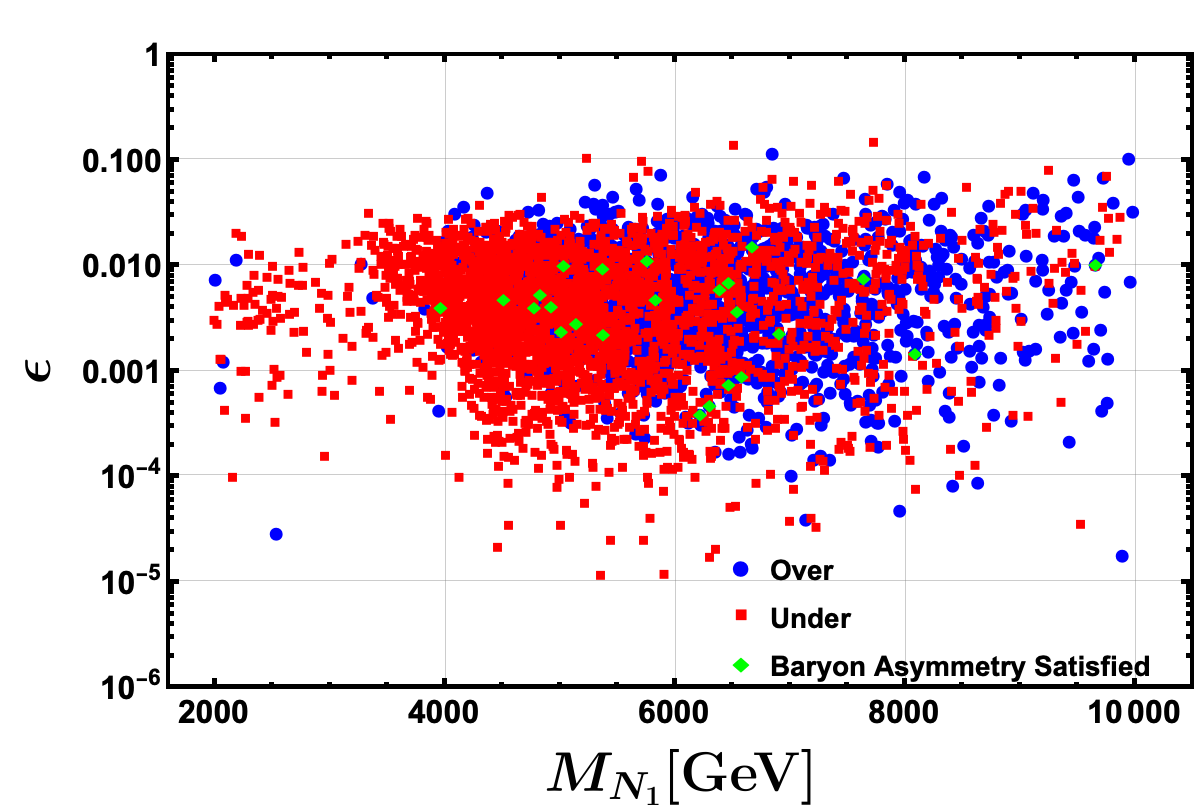}
\caption{Parameter scan in the $(M_{N_1},\,\epsilon)$ plane showing the regions corresponding to over-abundant (blue) and under-abundant (red) DM relic density. The green points satisfy the observed BAU. Here $M_{N_1}$ denotes the mass of the lightest heavy neutrino, while $\epsilon$ represents the CP-asymmetry parameter relevant for leptogenesis. \label{fig:eps_plot}}
\end{figure}

\begin{figure}[t]
\centering
\includegraphics[width=0.55\textwidth]{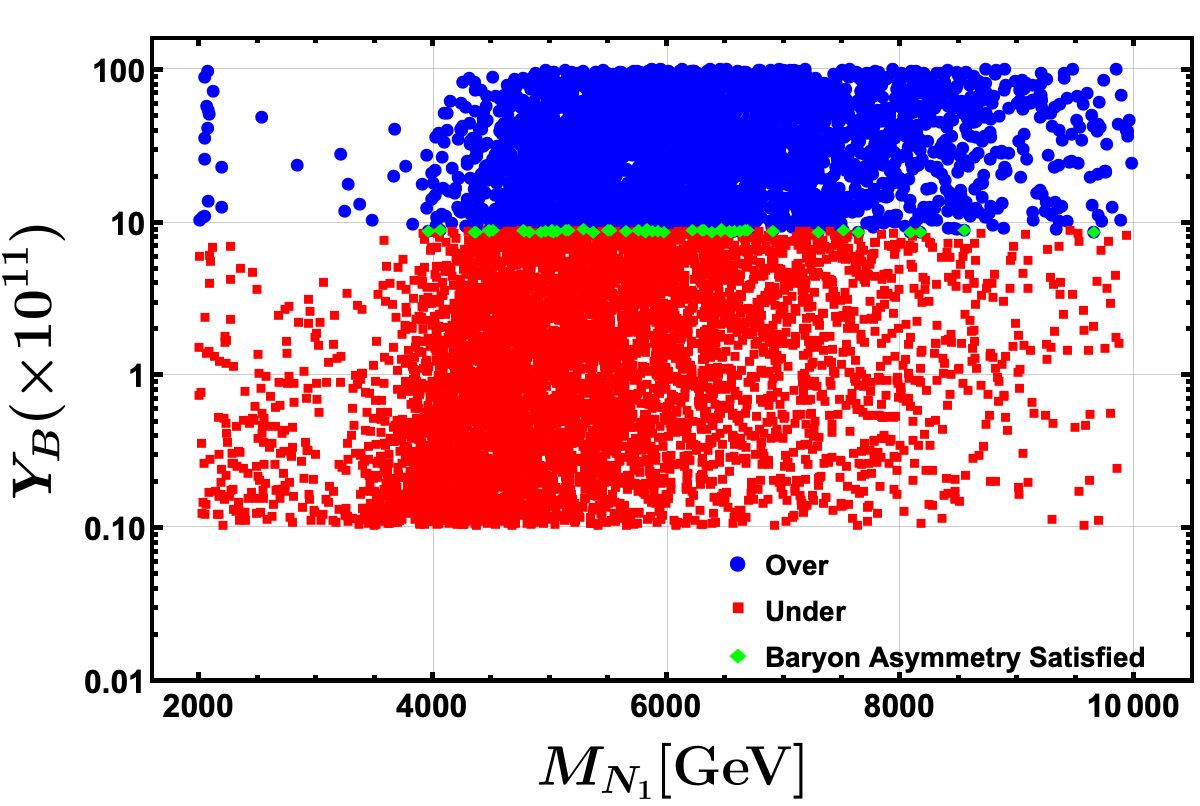}
\caption{Baryon asymmetry yield $Y_B$ as a function of the lightest heavy-neutrino mass $M_{N_1}$. The blue points correspond to parameter regions leading to over-abundant DM relic density, while the red points denote under-abundant relic-density regions. The green points satisfy the observed BAU and all constraints from DM, neutrino oscillations, direct detection, and LFV processes. \label{fig:yb_plot}}
\end{figure}

The outcome of the full parameter scan in the leptogenesis sector is illustrated in Figs.~\ref{fig:eps_plot} and \ref{fig:yb_plot}. First, Fig.~\ref{fig:eps_plot} shows the parameter dependence of the CP asymmetry $\varepsilon_1$ generated in the decay of $N_1$. A narrow resonant band appears when $M_{N_1}$ and $M_{N_2}$ are quasi-degenerate, where $\varepsilon_1$ is significantly enhanced, in agreement with the expectations of resonant leptogenesis. Away from this quasi-degenerate region, the CP asymmetry rapidly decreases, and successful leptogenesis becomes impossible. 

The resulting baryon asymmetry yield $Y_B$ obtained from solving the full Boltzmann equations for each scanned point is presented in Fig.~\ref{fig:yb_plot}. The green points in this plot denote the subset of points that \emph{simultaneously} reproduce the observed baryon asymmetry and satisfy all phenomenological constraints, namely the DM relic density, neutrino oscillation data, direct-detection limits, and the LFV bounds on $\ell_i \to \ell_j \gamma$. These green points are therefore identified as the fully allowed region of the model.

\begin{figure}[t]
\centering
\includegraphics[width=0.47\textwidth]{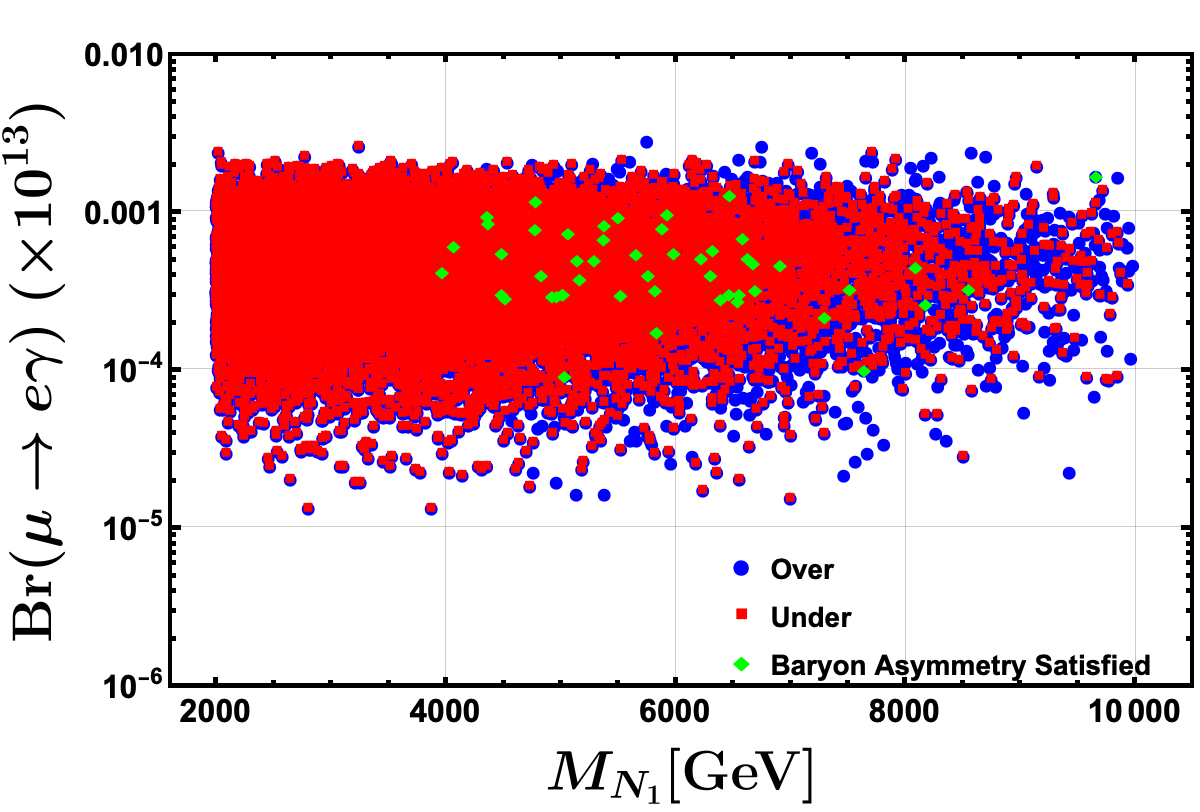}\\[0.5cm]
\includegraphics[width=0.47\textwidth]{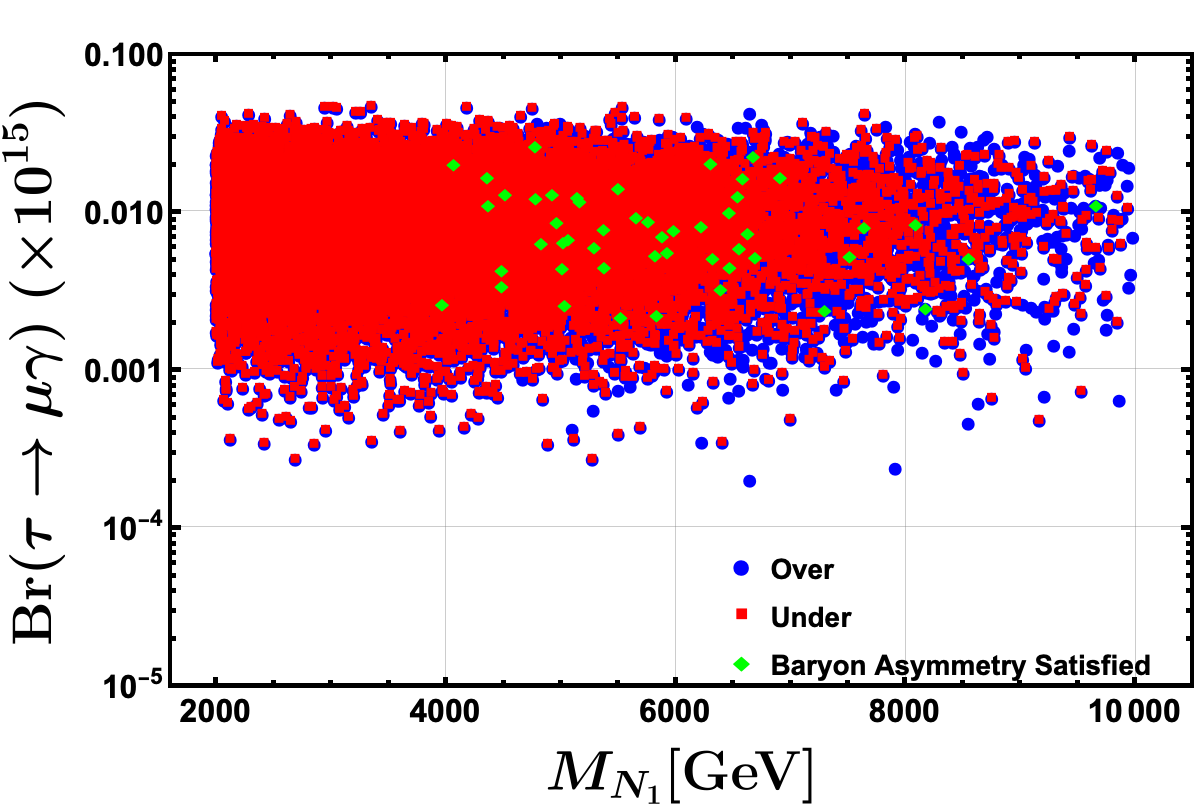}\hfill
\includegraphics[width=0.47\textwidth]{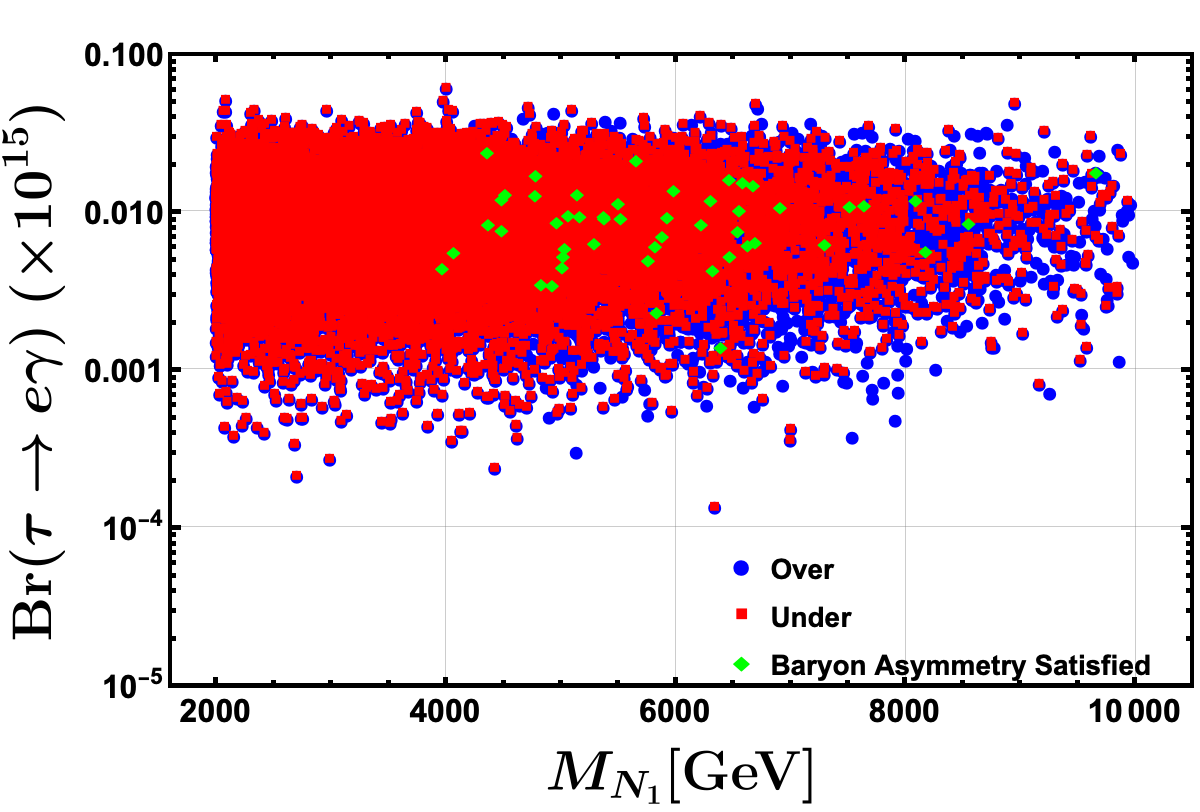}
\caption{Predicted branching ratios of the radiative LFV processes as functions of the lightest heavy neutrino mass $M_{N_1}$. The blue and red points correspond to parameter regions yielding over-abundant and under-abundant DM relic density respectively, while the green points satisfy the observed BAU. \label{fig:BR_N1}}
\end{figure}

The complementary impact of the LFV constraints is shown in Fig.~\ref{fig:BR_N1}, which displays the predicted branching ratios for the radiative LFV decays $\ell_i \to \ell_j \gamma$ within the same parameter space. Each point corresponds to a fully numerical evaluation of the loop-induced amplitude for the chosen set of masses and Yukawa couplings. The plots demonstrate that the green points in Fig.~\ref{fig:yb_plot} lie comfortably below the current experimental upper limits on these LFV processes, while the regions outside tend to generate larger LFV rates and are more strongly constrained.

From the allowed green points, we select two representative benchmark points, which satisfy all the phenomenological constraints discussed above: reproducing the observed neutrino oscillation pattern, achieving the correct DM relic abundance through $Z^\prime$-mediated annihilation, generating the observed baryon asymmetry through resonant leptogenesis, and respecting the bounds from $\ell_i \to \ell_j \gamma$. In both cases, the mass of the lightest heavy neutrino satisfies
\begin{equation}
M_{N_1} \sim \mathcal{O}(1\text{--}10\,\text{TeV}),
\end{equation}
which keeps the scenario experimentally testable, since such states can be produced at future high-energy colliders, in particular at multi-TeV muon colliders.

In summary, the interplay between the $Z^\prime$-mediated annihilation of DM, the suppressed scalar mixing, the quasi-degenerate heavy-neutrino spectrum, and the LFV constraints yields a narrow yet phenomenologically viable region of parameter space. Within this domain, both the DM relic density and the observed baryon asymmetry can be simultaneously reproduced without conflicting with collider, direct-detection, or flavor data. Representative benchmark points that satisfy all the experimental and cosmological requirements are listed in Table~\ref{tab:benchmarks}.

\begin{table}[t]
\centering
\renewcommand{\arraystretch}{1.3}
\setlength{\tabcolsep}{15pt}
\caption{Benchmark parameter sets consistent with the observed DM relic abundance, baryon asymmetry, and recent LZ-2024 direct detection limits. The relation $m_\chi \simeq 2 m_{N_1}$ and small scalar mixing $\theta \sim \mathcal{O}(0.01)$ are crucial for ensuring compatibility with all the current experimental and cosmological observations. \label{tab:benchmarks}}
\vspace{0.2cm}
\begin{tabular}{lcc}
\hline\hline
\textbf{Parameter} 
& \textbf{Benchmark 1 (BP1)} 
& \textbf{Benchmark 2 (BP2)} 
 
\\
\hline
$m_{Z^\prime}~[\text{GeV}]$              & 7000 & 7000  \\
$m_{\rho} \equiv M_{H_2}~[\text{GeV}]$ & 500  & 500    \\
$m_{\chi}~[\text{GeV}]$             & 8200 & 8737  \\
$M_{N_1}~[\text{GeV}]$              & 4100 & 4368 \\
$M_{N_2}~[\text{GeV}]$              & 4100 & 4368  \\
$v_{\phi}~[\text{GeV}]$             & 5000 & 5000 \\
$\theta$                            & $\mathcal{O}(0.01)$ & $\mathcal{O}(0.01)$  \\
\hline\hline
\end{tabular}
\end{table}

\section{Collider Phenomenology}
\label{sec:collider}

The parameter space selected in the preceding section, where the DM relic abundance, the direct- and indirect-detection limits, and the $B-L$ asymmetry are simultaneously satisfied, naturally points to a TeV-scale spectrum for the new states. 

At hadron colliders, the Drell–Yan production of $Z^\prime$ followed by $Z^\prime\!\to \ell^+\ell^-$ provides the leading probe of the gauge sector. Present dilepton searches push $m_{Z^\prime}$ to several TeV, which is well accommodated in the chosen benchmarks. Direct production of the heavy neutrinos at the LHC suffers from small cross sections and substantial SM backgrounds, making discovery challenging with cut-based selections. A lepton collider with centre‑of‑mass energies in the few to 10 TeV range and clean experimental conditions offers substantially improved sensitivity to TeV‑scale heavy neutrinos through active–sterile mixing. A future muon collider, in particular, would be ideally suited to realise this potential.

At a muon collider, the dominant production mechanism for heavy Majorana neutrinos is the single-production process
\begin{equation}
    \mu^+\mu^- \;\to\; N_i\,\nu_\ell \qquad (\ell=e,\mu,\tau),
    \label{eq:singleN}
\end{equation}
mediated primarily by the $t$-channel $W$ exchange, with the sub-leading $s$-channel $Z$ contributions controlled by the active–sterile mixing. The produced heavy neutrino subsequently decays via
\begin{equation}
N_i \to W^\pm \ell^\mp,\qquad N_i \to Z\nu_\ell,\qquad N_i \to h\nu_\ell,
\end{equation}
leading to two characteristic final-state signatures: a fully leptonic dilepton mode, $2\ell + \slashed{E}_T$, arising from $W^\pm\!\to \ell^\pm \nu_\ell$, and a semi-leptonic mode, $1\ell + 2j + \slashed{E}_T$, resulting from the hadronic decay $W^\pm\!\to jj$. These two channels provide complementary sensitivity to the heavy-neutrino dynamics. 

Before discussing these channels, we make an important comment. A major experimental challenge at high-energy muon colliders is the presence of a beam-induced background (BIB), originating from the decay of muons in the circulating beams~\cite{Homiller:2022iax,Ally:2022rgk,Antonelli:2015nla}. These decays generate a large flux of predominantly soft, low-energy particles that can significantly affect detector performance if not properly mitigated. Dedicated mitigation strategies, such as shielding nozzles in the forward region and high-precision timing in tracking and calorimeter systems, have been shown to substantially suppress this background. For the signal processes considered in this work, involving heavy neutrinos with masses of order $\mathcal{O}(4\,\mathrm{TeV})$, the resulting final states are characterized by highly energetic charged leptons and, in the semi-leptonic channel, hard jets. The transverse momenta of these objects are significantly larger than the typical energy scale associated with BIB particles. Imposing a selection requirement of $p_T > 10\,\mathrm{GeV}$ for charged leptons within the detector acceptance $|\eta| < 2.5$ efficiently suppresses the dominant BIB contributions. Therefore, for the benchmark scenarios considered in this analysis, the BIB impact is expected to be subdominant and does not significantly affect the signal sensitivity.

\subsection{Dilepton Final State \texorpdfstring{$2\ell + \slashed{E}_T$}{2lplusET}}

A particularly clean final state arises when the heavy neutrino decays through  
\begin{equation}
N_i \to W^\pm \ell^\mp,\qquad W^\pm \to \ell^\pm\nu_\ell,
\end{equation}
leading to the process
\begin{equation}
\mu^+\mu^- \to N_i\nu_\ell \to (\ell W)\,\nu_\ell \to \ell^+\ell^- + \slashed{E}_T.
\end{equation}
The dominant SM backgrounds include $\mu^+\mu^-\!\to W^+W^-\!\to \ell^+\ell^-+\slashed{E}_T$, off-shell $Z/\gamma^*\to \ell^+\ell^-$ in association with neutrinos, and multi-boson processes with leptonic decays. The analysis strategy is therefore based on basic acceptance cuts, hardness requirements on the lepton transverse momenta, a veto around the $Z$ pole in $m_{\ell\ell}$ to suppress $Z$-mediated backgrounds, and cuts on $\slashed{E}_T$ and angular variables (such as $\Delta\phi(\ell\ell)$) to reduce $WW$- and $ZZ$-induced contributions.

The signal and background samples are produced at leading order (LO) with the aid of \texttt{MG5aMC@NLO}~\cite{Alwall:2014hca}. Parton showering and hadronization are performed with \texttt{Pythia8}~\cite{Sjostrand:2014zea,Bierlich:2022pfr}, and the detector response is modeled using \texttt{Delphes}~\cite{deFavereau:2013fsa}, with a configuration appropriate for the proposed muon-collider. The jets are reconstructed with a collider-appropriate algorithm (e.g. the Valencia (VLC)~\cite{Boronat:2016tgd,Boronat:2014hva} algorithm) with radius $R=0.5$. Unless stated otherwise, the nominal object selections adopt
\begin{equation}
\begin{aligned}
p_{Tj} &> 20~\text{GeV}, \quad &|\eta_j| &< 5.0,\\[0.1cm]
p_{T\ell} &> 10~\text{GeV}, \quad &|\eta_\ell| &< 2.5,\\[0.1cm]
\Delta R_{ij} &> 0.4, \quad &i,j &= \{\ell,\,\text{jet}\},
\end{aligned}
\end{equation}
with $\Delta R_{ij}=\sqrt{(\eta_i-\eta_j)^2+(\phi_i-\phi_j)^2}$.

\begin{figure}[t]
    \centering
    \includegraphics[width=0.45\linewidth]{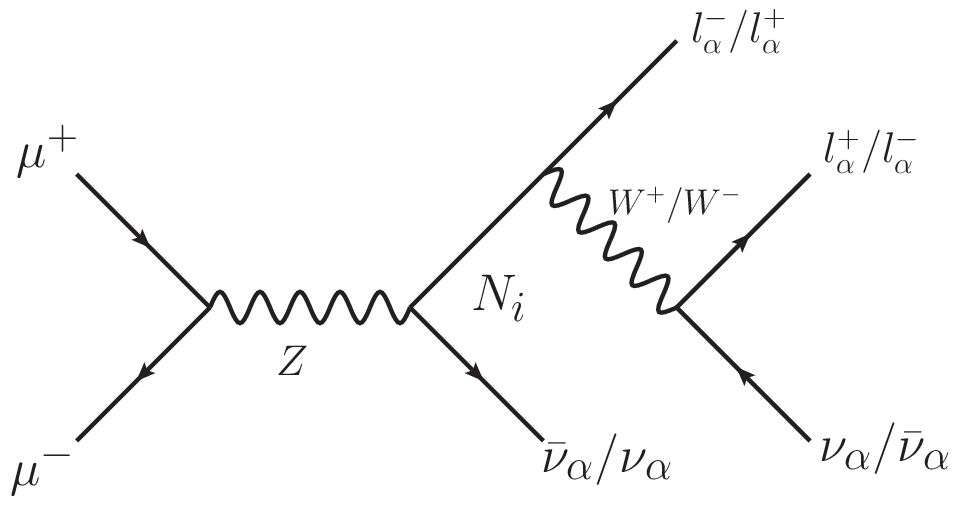}
    \includegraphics[width=0.45\linewidth]{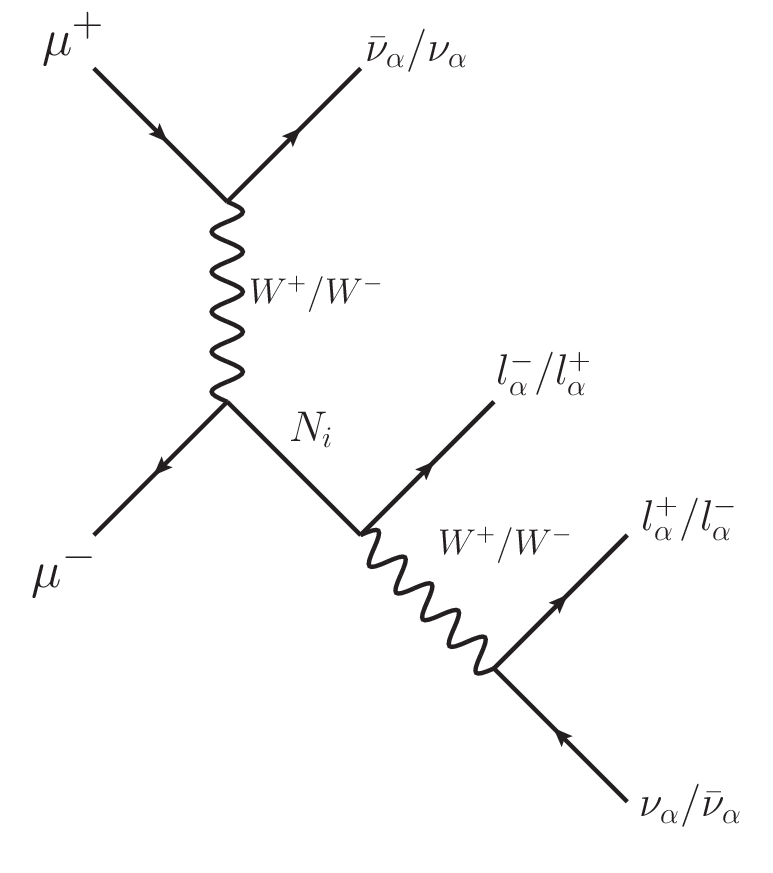}
    \caption{Representative Feynman diagrams for single heavy-neutrino production at a muon collider, $\mu^+\mu^- \to N_i\nu_\ell$, followed by $N_i \to W^\pm \ell^\mp$ and $W^\pm \to \ell^\pm \nu_\ell$, leading to the $\ell^+\ell^- + \slashed{E}_T$ final state.
    }
    \label{fig:feysig1}
\end{figure}

We focus on a final state characterized by two opposite-sign leptons, having the same or different flavor, and missing transverse energy, $\ell^+ \ell^- + \slashed{E}_T$, originating from
\begin{equation}
\mu^+ \mu^- \to N\,\nu_\ell, \quad N \to W^{\pm} \ell^{\mp}, \quad W^{\pm} \to \ell^{\pm} \nu_\ell(\bar{\nu}_\ell),
\end{equation}
where the heavy neutrino decays to a $W^\pm$ and a charged lepton through charged-current interactions, followed by the leptonic decay of $W^\pm$. Representative Feynman diagrams for the signal process are shown in Fig.~\ref{fig:feysig1}. The primary SM background mimicking this final state is $\mu^+ \mu^- \to \ell^+ \ell^- + \slashed{E}_T$, with dominant contributions from the following sub-processes:
\begin{itemize}
    \item $\mu^+ \mu^- \to \ell^- \ell^+ \to \ell^- W^{+*} \bar{\nu}_\ell \to \ell^- \ell^+ \nu_\ell \bar{\nu}_\ell$,
    \item $\mu^+ \mu^- \to \ell^- \ell^+ \to \ell^- Z^* \ell^+ \to \ell^- \ell^+ \nu_\ell \bar{\nu}_\ell$,
    \item $\mu^+ \mu^- \to \nu_i^* \nu_\ell \to \nu_\ell Z^* \bar{\nu}_\ell \to \nu_\ell \bar{\nu}_\ell \ell^+ \ell^-$,
    \item $\mu^+ \mu^- \to \nu_i^* \bar{\nu}_\ell \to \bar{\nu}_\ell W^* \ell^- \to \nu_\ell \bar{\nu}_\ell \ell^+ \ell^-$,
    \item $\mu^+ \mu^- \to W^+W^-,\, ZZ,\, W^+W^-Z,\, ZZZ$ with leptonic and semi-leptonic decays.
\end{itemize}
The LO cross sections for the process $\mu^+ \mu^- \to 2\ell + \slashed{E}_T$ at $\sqrt{s}=6$ and $10~\text{TeV}$ for the signal benchmarks and the total SM background are given in Table~\ref{tab:xsec}.

\begin{table}[t]
\centering
\renewcommand{\arraystretch}{1.3}
\setlength{\tabcolsep}{10pt}
\caption{LO cross sections for the process $\mu^+ \mu^- \to 2\ell + \slashed{E}_T$ at $\sqrt{s}=6$ and $10~\text{TeV}$ for the signal benchmarks and the SM background.}
\label{tab:xsec}
\vspace{0.5em}
\begin{tabular}{lcc}
\hline \hline
\textbf{Benchmark} & 
$\boldsymbol{\sigma_{\mathrm{LO}}}$ \textbf{(fb) @ $\sqrt{s}=6~\mathrm{TeV}$} &
$\boldsymbol{\sigma_{\mathrm{LO}}}$ \textbf{(fb) @ $\sqrt{s}=10~\mathrm{TeV}$} \\
\hline
BP1 Signal & 25  & 373 \\
BP2 Signal & 20  & 285 \\
Background & 233 & 198 \\
\hline \hline
\end{tabular}
\end{table}

\begin{figure}[t]
    \centering
    \includegraphics[width=0.49\textwidth]{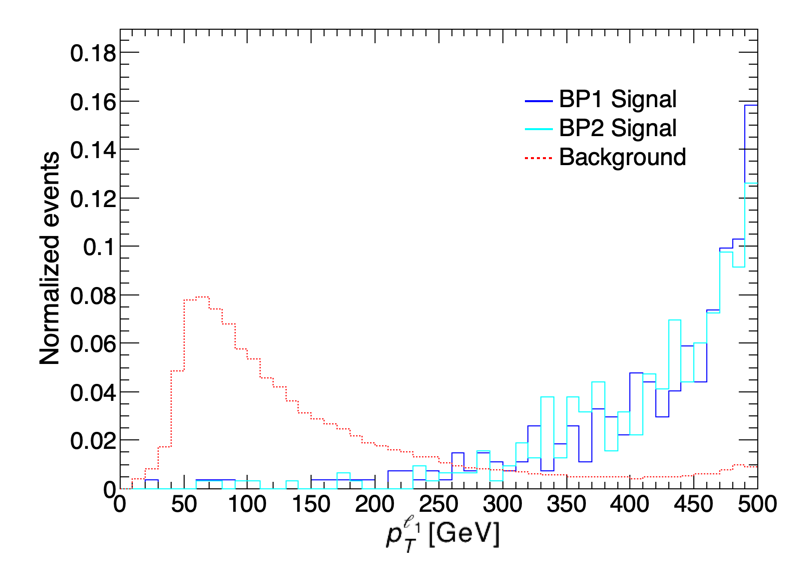}\hfill
    \includegraphics[width=0.49\textwidth]{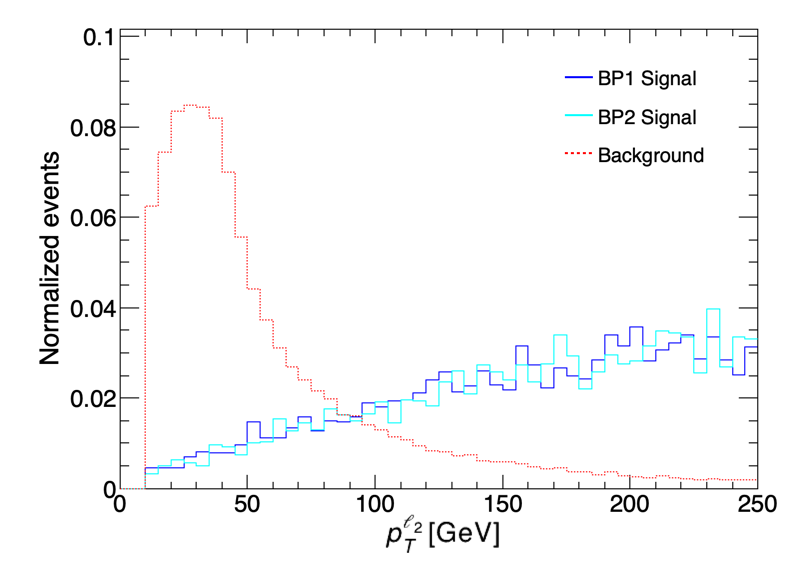}\\[0.3cm]
    \includegraphics[width=0.49\textwidth]{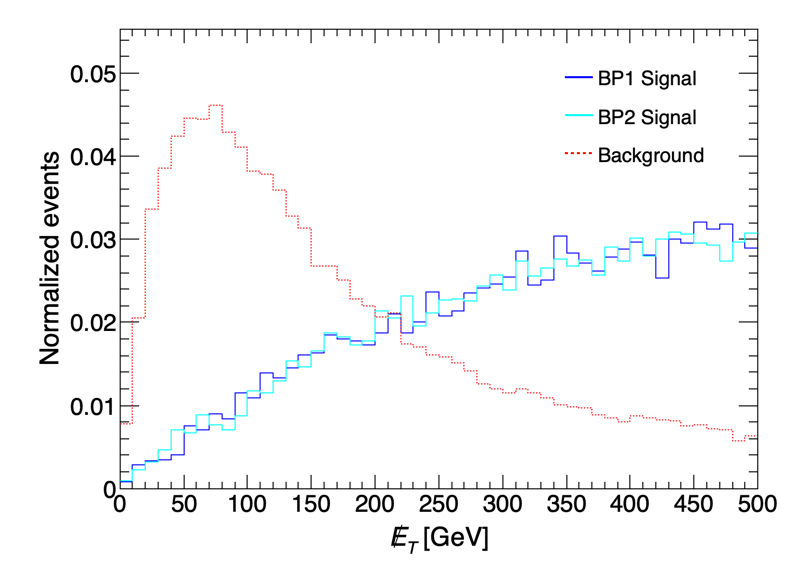}
    \caption{Normalized distributions of key kinematic observables for the process $\mu^+\mu^- \to 2\ell + \slashed{E}_T$ at $\sqrt{s} =6~\text{TeV}$: leading lepton transverse momentum $p_T^{\ell_1}$ (top-left), sub-leading lepton transverse momentum $p_T^{\ell_2}$ (top-right), and missing transverse energy $\slashed{E}_T$ (bottom). The curves correspond to the two signal benchmarks (BP1 and BP2) and the total SM background. These distributions are used to construct the optimized kinematic selections. \label{fig:kinematics}}
\end{figure}

\begin{figure}[t]
    \centering
    \includegraphics[width=0.49\textwidth]{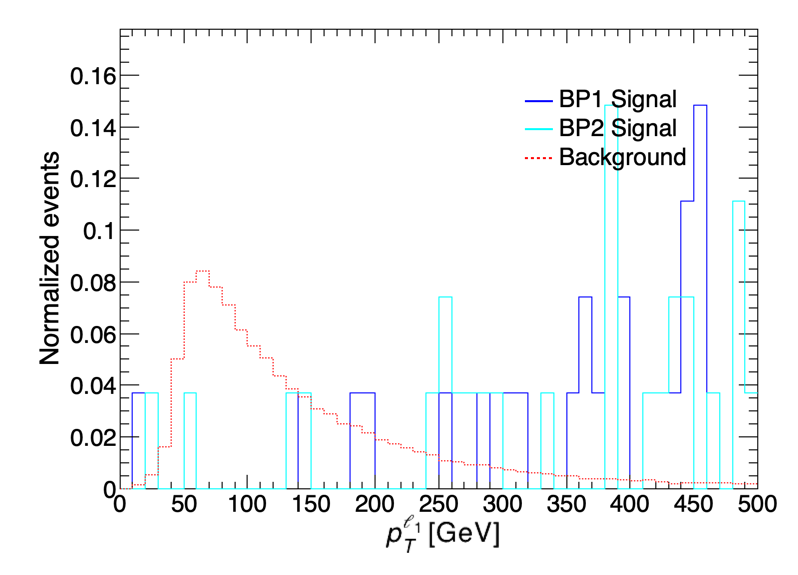}\hfill
    \includegraphics[width=0.49\textwidth]{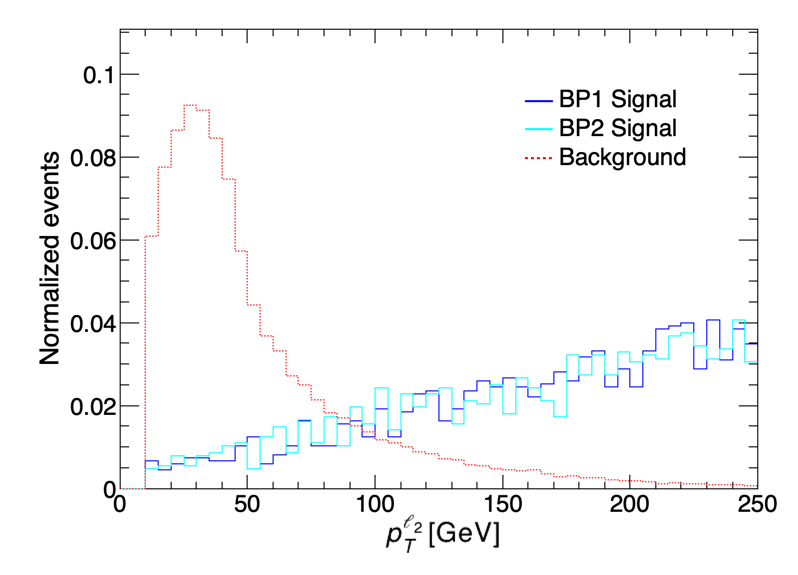}\\[0.3cm]
    \includegraphics[width=0.49\textwidth]{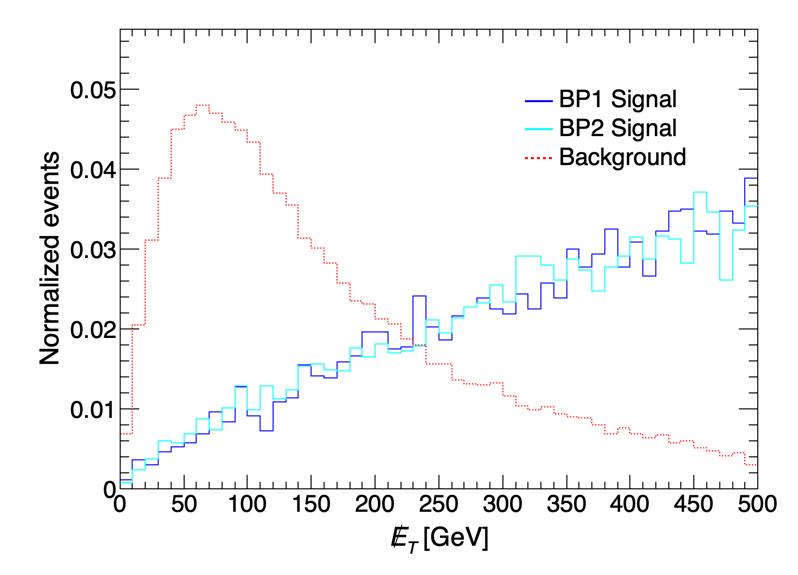}
    \caption{Same as in Fig.~\ref{fig:kinematics} but at $\sqrt{s}=10~\text{TeV}$. \label{fig:kinematics10tev}}
\end{figure}

To design an efficient set of analysis cuts, we first examine the shapes of several kinematic observables for the signal and the SM
background. From this exercise, three distributions are found to provide the most effective discrimination between signal and background: the transverse momenta of the leading and sub-leading leptons, $p_T^{\ell_1}$ and $p_T^{\ell_2}$, and the missing transverse energy $\slashed{E}_T$. Figs.~\ref{fig:kinematics} and \ref{fig:kinematics10tev} show the corresponding normalized distributions for the two signal benchmarks (BP1 and BP2) and the SM background at $\sqrt{s}=6$ and $10~\text{TeV}$, respectively. All distributions are normalized to unity to emphasize shape differences rather than absolute event rates. 

From Figs.~\ref{fig:kinematics} and \ref{fig:kinematics10tev}, we can see that the $p_T^{\ell_1}$ distribution (top-left panel) of the leading lepton in the signal is typically much harder than in the SM background, reflecting the large mass scale of the heavy neutrino produced. A similar, though somewhat milder, behavior is observed for $p_T^{\ell_2}$ (top-right panel), where the sub-leading lepton originating from the $W$ decay remains significantly harder in the signal than in the background. The missing transverse energy distribution (bottom panel) also exhibits a characteristic separation between signal and background, thereby providing an additional handle to suppress background events while retaining most of the signal. The same qualitative features observed at both center-of-mass energies motivate the use of an identical set of kinematic selection criteria in the subsequent analysis.

Guided by these kinematic distributions, we define the following sequence of selection cuts, denoted by $B_i$ for the $\sqrt{s}=6~\text{TeV}$ analysis:
\begin{itemize}
    \item \textbf{\boldmath $B_1$ (dilepton preselection):} we require exactly two isolated, oppositely charged leptons (electrons or muons) in the final state, and satisfying the basic acceptance criteria
    \begin{equation*}
    p_{T^\ell} > 10~\text{GeV}, \qquad |\eta_\ell| < 2.5, \qquad \Delta R_{\ell\ell} > 0.4.
    \end{equation*}

    \item \textbf{\boldmath $B_2$ (hard leading lepton):} taking advantage of the harder $p_T^{\ell_1}$ spectrum of the signal compared to the background (top-left panel of Fig.~\ref{fig:kinematics}), we impose
    \begin{equation*}
    p_T^{\ell_1} > 250~\text{GeV},
    \end{equation*}
    which removes a large fraction of the SM background while leaving the bulk of the signal events unaffected.

    \item \textbf{\boldmath $B_3$ (hard sub-leading lepton):} as can be seen from the top-right panel of Fig.~\ref{fig:kinematics}, the sub-leading lepton is also significantly harder in the signal. We therefore demand
    \begin{equation*}
    p_T^{\ell_2} > 100~\text{GeV},
    \end{equation*}
    efficiently suppressing the background configurations in which the second lepton is soft.

    \item \textbf{\boldmath $B_4$ (missing-energy selection):} guided by the shapes in the bottom panel of Fig.~\ref{fig:kinematics}, we apply a cut on the missing transverse energy,
    \begin{equation*}
    \slashed{E}_T < 250~\text{GeV},
    \end{equation*}
    which retains most of the signal while eliminating a substantial portion of the high-$\slashed{E}_T$ background tail.
\end{itemize}
The impact of these selections on the signal and background event yields is summarized in the cut-flow presented in Table~\ref{tab:cutflow}. After applying the complete set of cuts $B_1$–$B_4$, both the signal-to-background ratio and the statistical significance are significantly enhanced. For the benchmark scenarios considered, the integrated luminosity required to reach the $5\sigma$ discovery threshold in the $\ell^+\ell^- + \slashed{E}_T$ channel is at the level of a few fb$^{-1}$ at $\sqrt{s}=6~\text{TeV}$, and is further reduced at $\sqrt{s}=10~\text{TeV}$ due to the increased signal production rate.

\begin{table}[t]
\centering
\renewcommand{\arraystretch}{1.3}
\setlength{\tabcolsep}{8pt}
\caption{Cut-flow for signal and background events after sequential cuts $B_1$--$B_4$ at $\sqrt{s}=6$ and $10~\text{TeV}$, together with the integrated luminosity $\mathcal{L}_{5\sigma}$ required to reach $5\sigma$ significance. \label{tab:cutflow}}
\vspace{0.5em}
\begin{tabular}{lcccc|c}
\hline \hline
\textbf{Benchmark} 
& $B_1$ & $B_2$ & $B_3$ & $B_4$ 
& $\mathcal{L}_{5\sigma}$ (fb$^{-1}$) \\
\hline
\multicolumn{6}{c}{\boldmath$\sqrt{s}=6~\mathrm{TeV}$} \\
\hline
SM Background & 19033 & 8695 & 4862 & 3870 & -- \\
BP1 Signal    & 1967  & 1966 & 1775 & 1263 & 4.5 \\
BP2 Signal    & 1570  & 1569 & 1543 & 1003 & 5.7 \\
\hline
\multicolumn{6}{c}{\boldmath$\sqrt{s}=10~\mathrm{TeV}$} \\
\hline
SM Background & 15858 & 5166 & 2796  & 2796  & -- \\
BP1 Signal    & 29017 & 29016 & 28853 & 28853 & 0.095 \\
BP2 Signal    & 22113 & 22112 & 21982 & 21982 & 0.128 \\
\hline \hline
\end{tabular}
\end{table}

\subsection{\texorpdfstring{$1\ell + 2j + \slashed{E}_T$}{1lplus2jplusET} Final State}

\begin{figure}[t]
    \centering
    \includegraphics[width=0.45\linewidth]{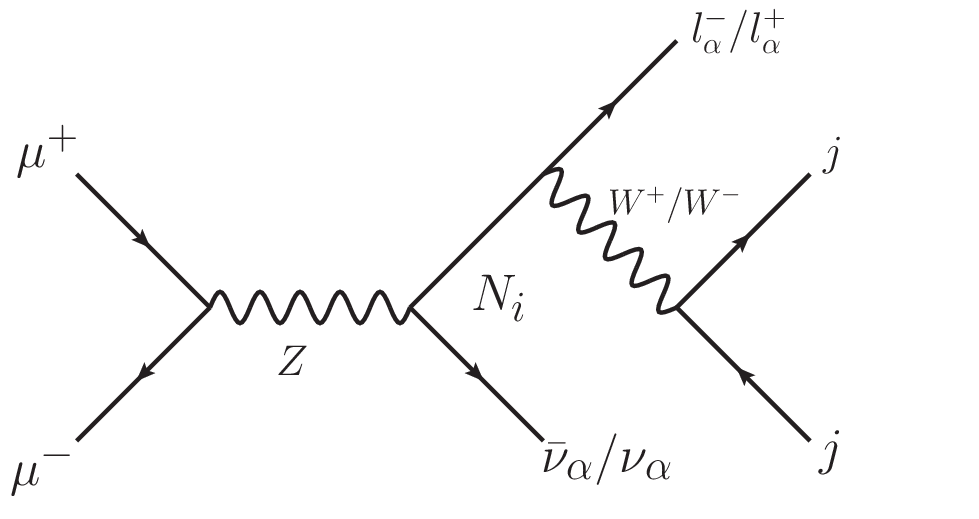}
    \includegraphics[width=0.45\linewidth]{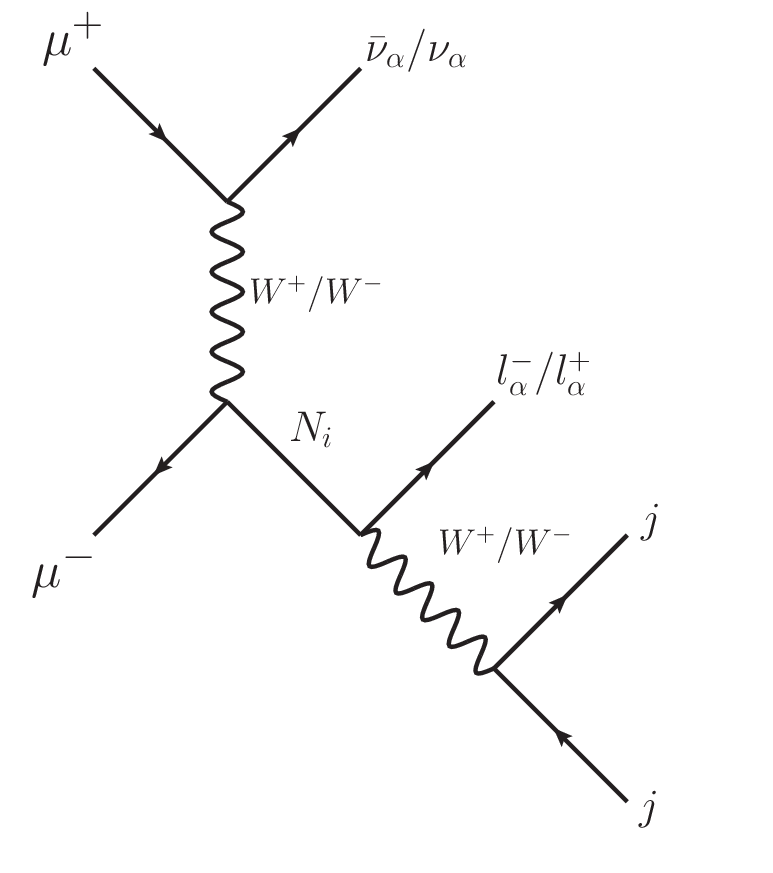}
    \caption{Representative Feynman diagrams for single heavy-neutrino production at a muon collider, $\mu^+\mu^- \to N_i\nu_\ell$, followed by $N_i \to W^\pm \ell^\mp$ and $W^\pm \to j j$, leading to the $\ell^\pm + 2j + \slashed{E}_T$ final state. \label{fig:feysig2}}
\end{figure}

A complementary and often more sensitive channel at a high-energy muon collider is the single-lepton plus dijet final state accompanied by missing transverse energy. In this topology, the heavy neutrino is produced in association with an active neutrino and subsequently undergoes a charged-current decay,
\begin{equation}
\mu^+ \mu^- \to \nu_\ell\, N_i,
\qquad
N_i \to W^\pm \ell^\mp,
\qquad
W^\pm \to jj,
\end{equation}
leading to the final state $\ell^\pm + 2j + \slashed{E}_T$. Representative Feynman diagrams for the signal process are shown in Fig.~\ref{fig:feysig2}. Compared to the fully leptonic mode, this channel benefits from the larger hadronic branching fraction of the $W$ boson and typically offers higher event rates.

The final-state particles are clustered into exactly two jets using a radius parameter of $R = 0.5$, ensuring stable reconstruction for multi-TeV kinematics. The jets are reconstructed using the VLC algorithm~\cite{Boronat:2016tgd,Boronat:2014hva}, which is well suited for the clean environment and boosted topologies at lepton colliders. The main SM processes capable of mimicking this signature are
\begin{itemize}
    \item $\mu^+\mu^- \to W^+W^- \to \ell^\pm \nu_\ell \; jj$,
    \item $\mu^+\mu^- \to ZZ$ with $Z\to jj$ and one lepton missed,
    \item $\mu^+\mu^- \to W^+W^-Z$ with partially leptonic $W$ decays,
    \item $\mu^+\mu^- \to ZZZ$ with $Z\to jj$ and invisible decays,
    \item $\mu^+\mu^- \to Zh$ with $h\to jj$ and a missed lepton.
\end{itemize}
After basic acceptance cuts, the $W^+W^-$ and $W^+W^-Z$ processes constitute the dominant irreducible backgrounds.

\begin{table}[t]
\centering
\renewcommand{\arraystretch}{1.3}
\setlength{\tabcolsep}{10pt}
\caption{LO cross sections for the process $\mu^+ \mu^- \to 1\ell + 2j + \slashed{E}_T$ at $\sqrt{s}=6$ and $10~\text{TeV}$.}
\label{tab:xsec-1l2j}
\vspace{0.5em}
\begin{tabular}{lcc}
\hline \hline
\textbf{Benchmark} & 
$\boldsymbol{\sigma_{\mathrm{LO}}}$ \textbf{(fb) at  $\sqrt{s}=6~\mathrm{TeV}$} &
$\boldsymbol{\sigma_{\mathrm{LO}}}$ \textbf{(fb) at $\sqrt{s}=10~\mathrm{TeV}$} \\
\hline
BP1 Signal   & 77                 & 1117 \\
BP2 Signal   & 67                 & 855  \\
$W^+W^-$     & 62.55              & 25.69 \\
$W^+W^-Z$    & 11.04              & 9.44  \\
$ZZ$         & $3.65\times10^{-2}$ &  $1.49\times10^{-2}$ \\
$Zh$         & 0.34             & 0.123 \\
$ZZZ$        & $4.93\times10^{-5}$ & $2.573\times10^{-5}$ \\
\hline \hline
\end{tabular}
\end{table}

Table~\ref{tab:xsec-1l2j} summarizes the LO cross sections for the signal benchmarks and the dominant SM backgrounds for the $1\ell + 2j + \slashed{E}_T$ final state at $\sqrt{s}=6$ and $10~\text{TeV}$. At $\sqrt{s}=6~\text{TeV}$, the signal rates are already sizable, with $\sigma_{\mathrm{BP1}} = 77~\text{fb}$ and $\sigma_{\mathrm{BP2}} = 67~\text{fb}$. These values originate from the enhanced single-production channel $\mu^+\mu^- \to N\nu_\ell$ mediated by $t$-channel $W$ exchange, which becomes increasingly efficient at multi-TeV center-of-mass energies and enables the probing of TeV-scale heavy neutrinos with appreciable production rates. Among the SM backgrounds, the $W^+W^-$ process constitutes the dominant contribution, with a cross section of $62.55~\text{fb}$ in the semi-leptonic topology relevant for the $1\ell + 2j + \slashed{E}_T$ final state. The next most significant background arises from the $W^+W^-Z$ production, yielding $11.04~\text{fb}$ once the decay modes with $Z\to jj$ and $Z\to\nu\bar{\nu}$ are included. Other electroweak channels, such as $ZZ$, $Zh$, and $ZZZ$, are strongly suppressed, contributing at the sub-femtobarn or even $10^{-2}~\text{fb}$ level, due to smaller electroweak couplings and the reduced phase space for multi-boson final states at $\sqrt{s}=6~\text{TeV}$.

At $\sqrt{s}=10~\text{TeV}$, the signal cross sections increase by more than an order of magnitude, while the dominant backgrounds exhibit a much weaker energy dependence. As a result, the signal-to-background hierarchy before selection cuts becomes even more favorable at higher energies. In general, the heavy-neutrino signal lies well above most SM backgrounds at the inclusive level and is primarily challenged by the
$W^+W^-$ and $W^+W^-Z$ channels.

\begin{figure}[t]
    \centering
    \includegraphics[width=0.49\textwidth]{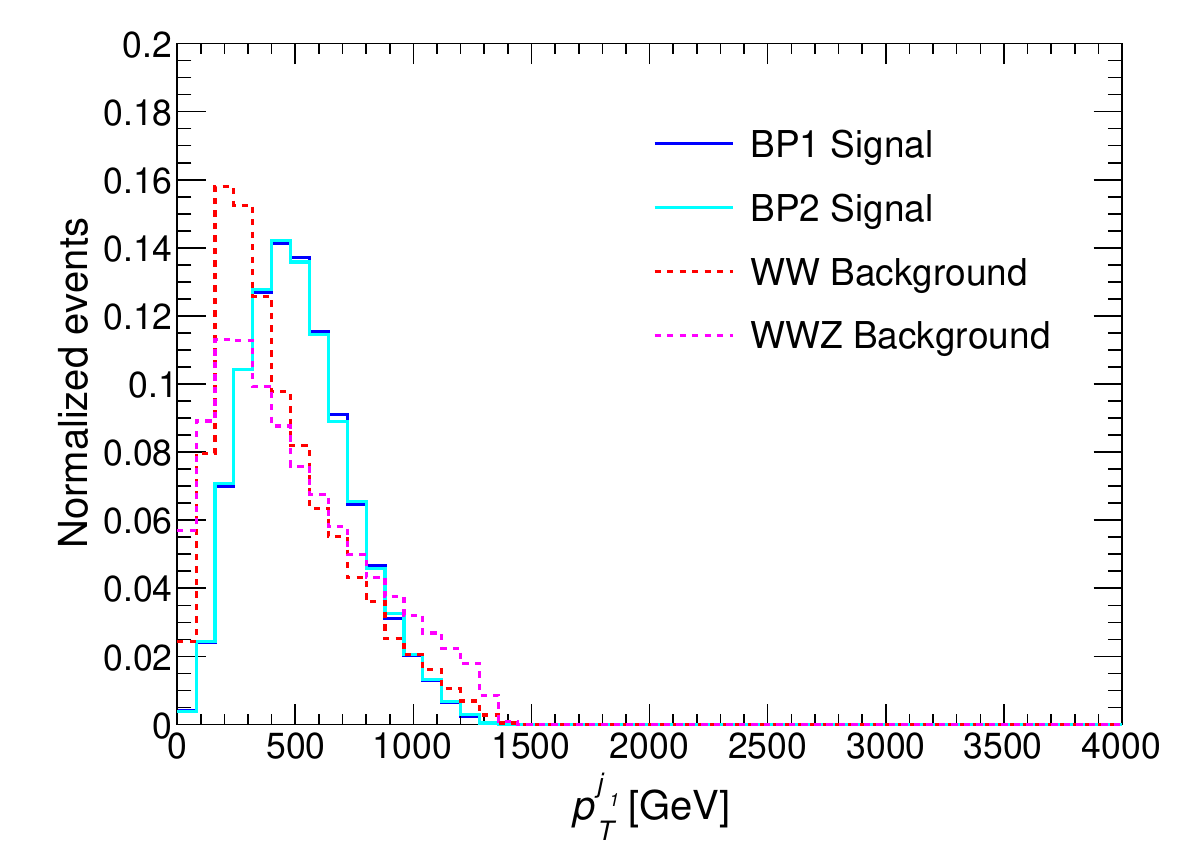}\hfill
    \includegraphics[width=0.50\textwidth]{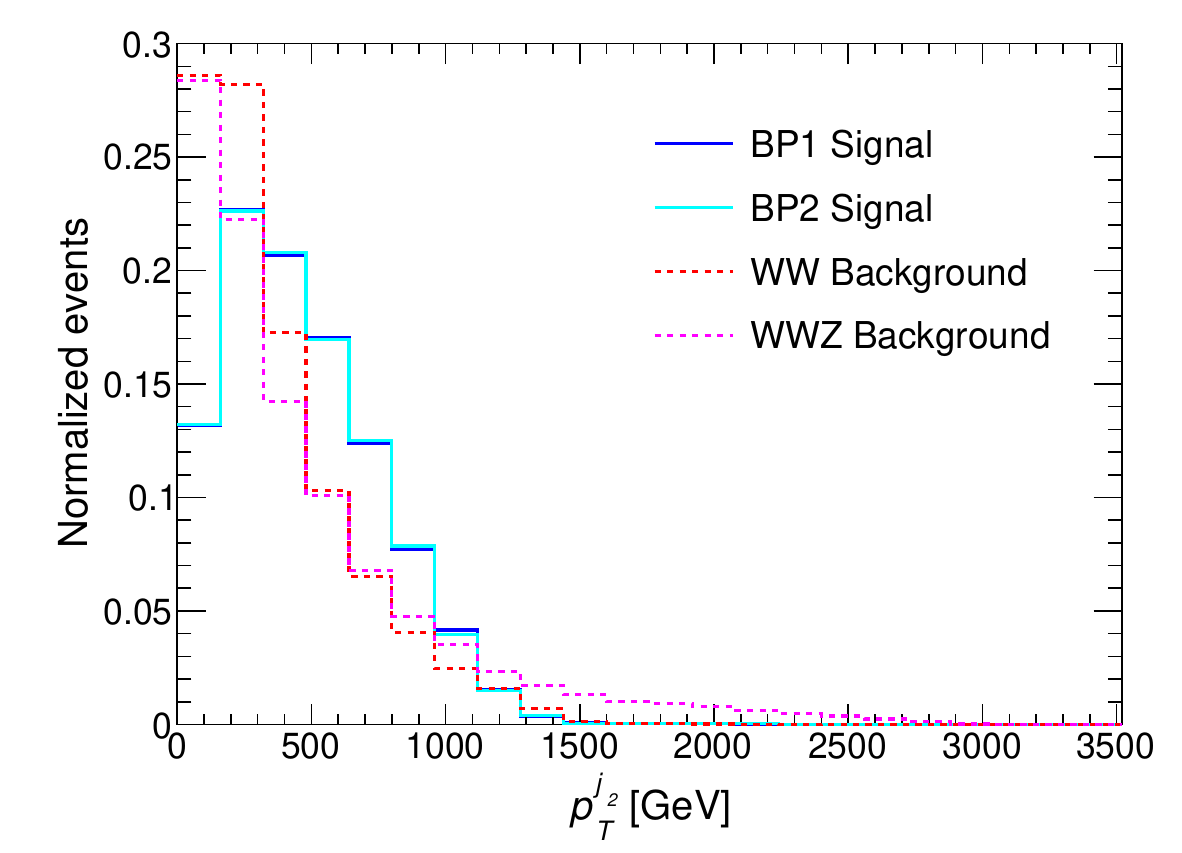}
    \caption{Normalized transverse momentum distributions of the jets for the $1\ell + 2j + \slashed{E}_T$ final state at $\sqrt{s}=6~\mathrm{TeV}$. Left: leading jet $p_T^{j_1}$. Right: sub-leading jet $p_T^{j_2}$. Signal benchmarks BP1 and BP2 exhibit significantly harder jet spectra than the $W^+W^-$ and $W^+W^-Z$ backgrounds, reflecting the multi-TeV mass of the parent heavy neutrino. These features motivate the high-$p_T$ selections adopted in the analysis. \label{fig:jet-pt}}
\end{figure}

\begin{figure}[t]
    \centering
    \includegraphics[width=0.49\textwidth]{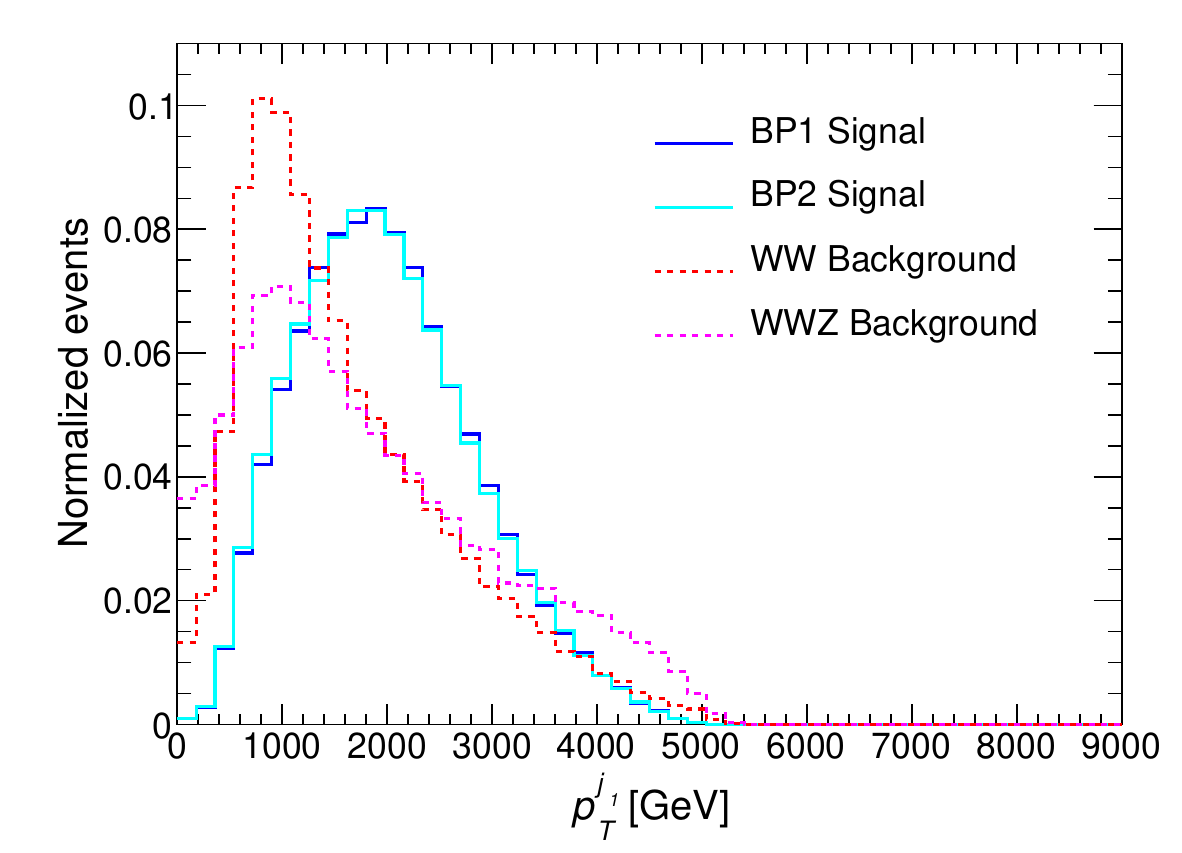}\hfill
    \includegraphics[width=0.49\textwidth]{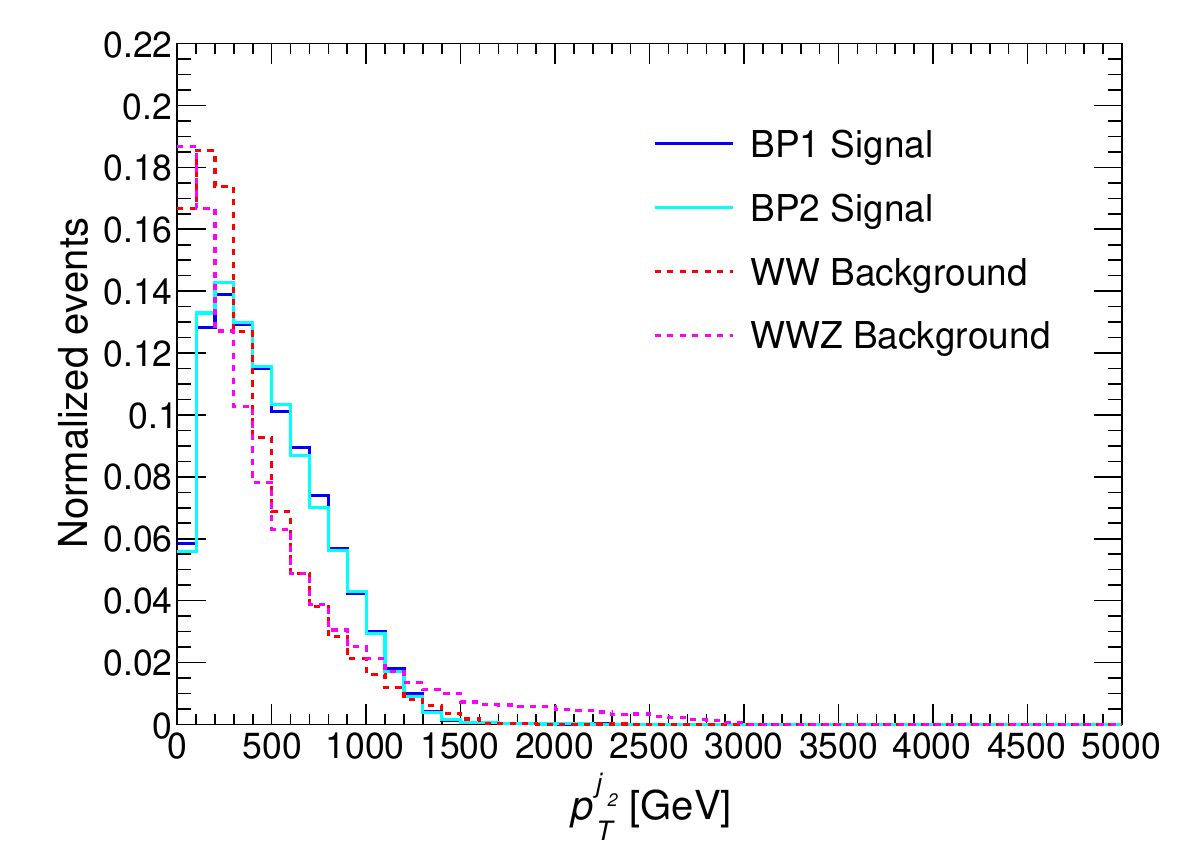}
    \caption{Same as in Fig.~\ref{fig:jet-pt} but at $\sqrt{s}=10~\mathrm{TeV}$. \label{fig:jet-pt10tev}}
\end{figure}

To construct an efficient set of selection criteria, we examine the shapes of key observables sensitive to the presence of a heavy neutrino.
Figures~\ref{fig:jet-pt} and \ref{fig:jet-pt10tev} show the normalized distributions of the transverse momenta of the leading and sub-leading jets, $p_T^{j_1}$ and $p_T^{j_2}$, for the signal benchmarks and the dominant backgrounds at $\sqrt{s}=6$ and $10~\text{TeV}$. In both cases, the signal distributions are characterized by substantially harder jet spectra than those of the background, reflecting the boosted hadronic $W$ boson produced in the decay of a multi-TeV heavy neutrino. The separation between signal and background is striking, based on the following observations:  
\begin{itemize}
    \item The leading jet from $W\to jj$ in heavy neutrino decays carries a large fraction of the heavy neutrino mass, producing a peak well above $1$--$2$~TeV for the two benchmarks.
    \item The sub-leading jet is also harder than that of the background, typically exceeding several hundred GeV.
    \item The background jets from $W^+W^-$ peak at significantly lower $p_T$, reflecting the SM $W$ mass scale.
\end{itemize}
These observations make the $1\ell + 2j + \slashed{E}_T$ channel a powerful discriminator by applying high-$p_T$ jet selections.

Motivated by the distributions shown in Figs.~\ref{fig:jet-pt} and \ref{fig:jet-pt10tev}, we adopt the following optimized requirements:
\begin{itemize}
    \item \textbf{Cut J1 (leading jet):}
    \begin{equation*}
    p_T^{j_1} > 800~\text{GeV}.
    \end{equation*}
    \item \textbf{Cut J2 (sub-leading jet):}
    \begin{equation*}
    p_T^{j_2} > 400~\text{GeV}.
    \end{equation*}
\end{itemize}
The numerical values shown in Table~\ref{tab:1l2j-cutflow} quantify the impact of the jet-based selections on the heavy-neutrino signal and the dominant electroweak backgrounds. All event yields are shown for a common reference with an integrated luminosity of $100~\text{fb}^{-1}$. At $\sqrt{s}=6~\text{TeV}$, the dominant background contributions arise from the $W^+W^-$ and $W^+W^-Z$ productions, yielding 6255 and 1104 events respectively, before the application of any analysis cuts. The corresponding signal yields for the benchmark points BP1 and BP2 are 7700 and 6100 events, indicating that the signal already competes with the leading backgrounds at the inclusive level.

\begin{table}[t]
\centering
\renewcommand{\arraystretch}{1.3}
\setlength{\tabcolsep}{10pt}
\caption{Cut-flow for the $1\ell + 2j + \slashed{E}_T$ final state at $\sqrt{s}=6$ and $10~\text{TeV}$, showing event yields after applying the leading-jet ($p_T^{j_1} > 800~\text{GeV}$) and sub-leading-jet ($p_T^{j_2} > 400~\text{GeV}$) selections. The integrated luminosity
$\mathcal{L}_{5\sigma}$ required to reach $5\sigma$ significance is also shown.}
\label{tab:1l2j-cutflow}
\vspace{0.5em}
\begin{tabular}{lcccc}
\hline\hline
\textbf{Benchmark} & Before cut & $J_1$ & $J_2$ & $\mathcal{L}_{5\sigma}$ (fb$^{-1}$) \\
\hline
\multicolumn{5}{c}{\boldmath$\sqrt{s}=6~\mathrm{TeV}$} \\
\hline
$W^{+}W^{-}$     & 6255 & 424  & 209  & -- \\
$W^{+}W^{-}Z$    & 1104 & 210  & 127  & -- \\
\hline
BP1 Signal       & 7700 & 4735 & 2827 & 0.88 \\
BP2 Signal       & 6100 & 3741 & 2229 & 1.02 \\
\hline
\multicolumn{5}{c}{\boldmath$\sqrt{s}=10~\mathrm{TeV}$} \\
\hline
$W^{+}W^{-}$     & 2565 & 295  & 233  & -- \\
$W^{+}W^{-}Z$    & 944 & 140  & 110  & -- \\
\hline
BP1 Signal       & 111700 & 91704 & 86101 &  0.029  \\
BP2 Signal       & 85500 & 70210 & 65802 &   0.038  \\
\hline\hline
\end{tabular}
\end{table}

Imposing the leading-jet requirement $p_T^{j_1} > 800~\text{GeV}$ (cut~$J_1$) results in a strong suppression of the backgrounds at
$\sqrt{s}=6~\text{TeV}$: the $W^+W^-$ yield is reduced from 6255 to 424 events, and the $W^+W^-Z$ contribution from 1104 to 210 events. In contrast, the signal retains a large fraction of events, with BP1 decreasing from 7700 to 4735 and BP2 from 6100 to 3741 events. This behavior directly reflects the characteristic hardness of the jets in the signal, which is largely absent in SM diboson production.

The additional requirement on the sub-leading jet, $p_T^{j_2} > 400~\text{GeV}$ (cut~$J_2$), further sharpens the discrimination. After this selection, the $W^+W^-$ background is reduced to 209 events, while the $W^+W^-Z$ background decreases to 127 events. The signal remains comparatively robust, with 2827 events for BP1 and 2229 events for BP2. In general, the combined jet selections suppress the dominant backgrounds by factors of $\mathcal{O}(30)$, while preserving approximately one third of the signal, in line with expectations for a boosted two-jet system arising from heavy-neutrino decay.

At $\sqrt{s}=10~\text{TeV}$, the inclusive signal yields increase dramatically, reaching 111700 events for BP1 and 85500 events for BP2 before
cuts, while the dominant backgrounds remain at the level of a few thousand events. As a result, the signal-to-background ratio is already highly favorable before any kinematic selection. The jet-based requirements are therefore not essential for discovery at this energy, but are instead introduced to enhance sample purity and improve the robustness of the analysis against detector effects and additional reducible backgrounds.

The integrated luminosity required to reach a $5\sigma$ discovery significance reflects this behavior. At $\sqrt{s}=6~\text{TeV}$, the values of $\mathcal{L}_{5\sigma} \simeq 0.88~\text{fb}^{-1}$ for BP1 and $\mathcal{L}_{5\sigma} \simeq 1.02~\text{fb}^{-1}$ for BP2 are sufficient
after the full set of selections. At $\sqrt{s}=10~\text{TeV}$, the required luminosity drops to the level of ${\cal O}(10^{-2})~\text{fb}^{-1}$, even with minimal or no jet-based cuts. These results highlight the excellent sensitivity of the $1\ell + 2j + \slashed{E}_T$ channel at a multi-TeV muon collider, and demonstrate that heavy-neutrino signals can be isolated with high statistical significance using simple and robust selection criteria.

In summary, because the heavy-neutrino production yields clean leptonic and semi-leptonic signatures, while complementary information on the $U(1)_{B-L}$ gauge sector can be extracted from precision dilepton measurements, a multi-TeV muon collider provides robust sensitivity to the neutrino-sector benchmarks identified in our cosmological analysis. Together with hadron-collider searches for $Z^\prime$ bosons and non-collider input such as the relic abundance and direct-detection limits, the collider reach presented here completes a coherent and comprehensive test of the parameter region motivated by the DM relic density and successful leptogenesis.

\section{Conclusion}
\label{sec:summary}

We have presented a predictive and experimentally testable framework in which the origin of neutrino masses, the BAU, and the nature of DM emerge from a correlated dynamical structure within a local $U(1)_{B-L}$ extension of the SM. A key distinguishing feature of this construction, relative to previous $B-L$ scenarios, is the implementation of an inverse-seesaw mechanism that involves the sterile fermion $S_{1}$ and the complex scalar field $\phi$. The VEV of $\phi$ generates the masses of the heavy neutrinos and of the Majorana DM particle $\chi$ through common Yukawa interactions, while a suppressed LNV parameter arising from a higher-dimensional operator yields naturally light active neutrinos together with TeV-scale heavy states that can be probed at colliders.

As the heavy neutrinos, the DM particle, and the $Z^\prime$ boson all acquire masses from the same $U(1)_{B-L}$ breaking sector, the early-Universe cosmology is tightly interconnected. A complete numerical solution of the coupled Boltzmann equations was carried out, tracking the evolution of $N_{1,2}$, $\chi$, and the net $B-L$ charge. In this setup, resonant leptogenesis becomes efficient for quasi-degenerate heavy-neutrino masses at the TeV scale. At the same time, the relic abundance of $\chi$ is dominantly shaped by annihilation through the $Z^\prime$ portal, with the correct DM density obtained with a small scalar-Higgs mixing angle that remains consistent with strong direct-detection limits.

A global parameter-space analysis incorporating neutrino oscillation data, radiative LFV processes, DM direct detection, collider constraints on $N_{1,2}$ and $Z^\prime$, and successful baryogenesis identifies a remarkably narrow but stable region in which all requirements are satisfied simultaneously. In this region, representative benchmarks are found with
\begin{equation*}
m_{Z^\prime} \simeq 7~\text{TeV}, \qquad
m_\chi \simeq 8\text{--}9~\text{TeV}, \qquad
M_{N_{1,2}} = \mathcal{O}(4\text{--}5)~\text{TeV}, \qquad
v_\phi \simeq 5~\text{TeV}. 
\end{equation*}
The inverse-seesaw structure allows light-neutrino masses to be generated with sizable active-sterile mixing and TeV-scale heavy neutrinos, making the sterile sector potentially testable at colliders. In the present setup, this feature is combined with resonant leptogenesis and the $U(1)_{B-L}$ portal to obtain a correlated explanation of neutrino masses, the BAU, and the DM.

The collider implications of this scenario are equally distinctive. Although LHC dilepton searches already constrain the $Z^\prime$ mass above the multi-TeV scale, the heavy neutrinos remain elusive at hadron colliders due to small active-sterile mixing and overwhelming backgrounds. In contrast, the clean environment and high-energy reach of future multi-TeV muon colliders provide a uniquely sensitive probe of this model. A detailed Monte-Carlo analysis was performed for both the fully leptonic $2\ell + \slashed{E}_T$ channel and the hadronic assisted $1\ell + 2j + \slashed{E}_T$ final state. By exploiting the hardness of leptons, boosted-jet kinematics, and large missing transverse energy, the dominant SM backgrounds can be reduced by more than an order of magnitude while retaining high signal efficiency. For the benchmarks satisfying all cosmological and low-energy constraints, a discovery at the level of $5\sigma$ is achievable with luminosities of order $\mathcal{O}(1)\,\text{fb}^{-1}$ at $\sqrt{s}=6$~TeV and at $\mathcal{O}(0.1)\,\text{fb}^{-1}$ at $\sqrt{s}=10$~TeV .

Taken together, the results highlight a coherent and highly constrained scenario in which the inverse-seesaw origin of neutrino masses, the observed baryon asymmetry, and the DM relic density stem all from a single gauge-extended structure. The preferred region of parameter space is not only cosmologically motivated but also within the reach of next-generation experimental programs. In particular, multi-TeV muon colliders, supported by improved LFV searches and DM experiments, offer a realistic possibility of decisively testing this class of models. 

We therefore conclude that TeV-scale $U(1)_{B-L}$ models with an inverse-seesaw structure can simultaneously address several fundamental open problems while remaining sharply predictive. The convergence of cosmological, astrophysical, and collider observables points toward a compelling direction for future exploration, with upcoming facilities poised to validate or exclude this scenario in a definitive manner.

\acknowledgments
This work is supported by the National Natural Science Foundation of China under Grant Nos.~12475094, 12135006, and 12575099, as well as the Science and Technology Innovation Leading Talent Support Program of Henan Province under Grant No.~254000510039. XY is also supported in part by the Startup Research Funding from CCNU.

\bibliographystyle{JHEP}
\bibliography{ref}

\providecommand{\href}[2]{#2}\begingroup\raggedright\begin{thebibliography}{100}

\bibitem{Aad:2012tfa}
{\scshape ATLAS} collaboration, \emph{{Observation of a new particle in the search for the Standard Model Higgs boson with the ATLAS detector at the LHC}}, \href{https://doi.org/10.1016/j.physletb.2012.08.020}{\emph{Phys. Lett. B} {\bfseries 716} (2012) 1} [\href{https://arxiv.org/abs/1207.7214}{{\ttfamily 1207.7214}}].

\bibitem{Chatrchyan:2012ufa}
{\scshape CMS} collaboration, \emph{{Observation of a New Boson at a Mass of 125 GeV with the CMS Experiment at the LHC}}, \href{https://doi.org/10.1016/j.physletb.2012.08.021}{\emph{Phys. Lett. B} {\bfseries 716} (2012) 30} [\href{https://arxiv.org/abs/1207.7235}{{\ttfamily 1207.7235}}].

\bibitem{Higgs:1964pj}
P.W.~Higgs, \emph{{Broken Symmetries and the Masses of Gauge Bosons}}, \href{https://doi.org/10.1103/PhysRevLett.13.508}{\emph{Phys. Rev. Lett.} {\bfseries 13} (1964) 508}.

\bibitem{Englert:1964et}
F.~Englert and R.~Brout, \emph{{Broken Symmetry and the Mass of Gauge Vector Mesons}}, \href{https://doi.org/10.1103/PhysRevLett.13.321}{\emph{Phys. Rev. Lett.} {\bfseries 13} (1964) 321}.

\bibitem{Guralnik:1964eu}
G.S.~Guralnik, C.R.~Hagen and T.W.B.~Kibble, \emph{{Global Conservation Laws and Massless Particles}}, \href{https://doi.org/10.1103/PhysRevLett.13.585}{\emph{Phys. Rev. Lett.} {\bfseries 13} (1964) 585}.

\bibitem{Higgs:1966ev}
P.W.~Higgs, \emph{{Spontaneous Symmetry Breakdown without Massless Bosons}}, \href{https://doi.org/10.1103/PhysRev.145.1156}{\emph{Phys. Rev.} {\bfseries 145} (1966) 1156}.

\bibitem{Super-Kamiokande:1998kpq}
{\scshape Super-Kamiokande} collaboration, \emph{{Evidence for oscillation of atmospheric neutrinos}}, \href{https://doi.org/10.1103/PhysRevLett.81.1562}{\emph{Phys. Rev. Lett.} {\bfseries 81} (1998) 1562} [\href{https://arxiv.org/abs/hep-ex/9807003}{{\ttfamily hep-ex/9807003}}].

\bibitem{SNO:2002tuh}
{\scshape SNO} collaboration, \emph{{Direct evidence for neutrino flavor transformation from neutral current interactions in the Sudbury Neutrino Observatory}}, \href{https://doi.org/10.1103/PhysRevLett.89.011301}{\emph{Phys. Rev. Lett.} {\bfseries 89} (2002) 011301} [\href{https://arxiv.org/abs/nucl-ex/0204008}{{\ttfamily nucl-ex/0204008}}].

\bibitem{Aghanim:2018eyx}
{\scshape Planck} collaboration, \emph{{Planck 2018 results. VI. Cosmological parameters}}, \href{https://doi.org/10.1051/0004-6361/201833910}{\emph{Astron. Astrophys.} {\bfseries 641} (2020) A6} [\href{https://arxiv.org/abs/1807.06209}{{\ttfamily 1807.06209}}].

\bibitem{Minkowski:1977sc}
P.~Minkowski, \emph{{$\mu \to e\gamma$ at a Rate of One Out of $10^{9}$ Muon Decays?}}, \href{https://doi.org/10.1016/0370-2693(77)90435-X}{\emph{Phys. Lett. B} {\bfseries 67} (1977) 421}.

\bibitem{Mohapatra:1979ia}
R.N.~Mohapatra and G.~Senjanovic, \emph{{Neutrino Mass and Spontaneous Parity Nonconservation}}, \href{https://doi.org/10.1103/PhysRevLett.44.912}{\emph{Phys. Rev. Lett.} {\bfseries 44} (1980) 912}.

\bibitem{Schechter:1980gr}
J.~Schechter and J.W.F.~Valle, \emph{{Neutrino Masses in SU(2) x U(1) Theories}}, \href{https://doi.org/10.1103/PhysRevD.22.2227}{\emph{Phys. Rev. D} {\bfseries 22} (1980) 2227}.

\bibitem{Schechter:1981cv}
J.~Schechter and J.W.F.~Valle, \emph{{Neutrino Decay and Spontaneous Violation of Lepton Number}}, \href{https://doi.org/10.1103/PhysRevD.25.774}{\emph{Phys. Rev. D} {\bfseries 25} (1982) 774}.

\bibitem{Mohapatra:2005wg}
R.N.~Mohapatra et~al., \emph{{Theory of Neutrinos: A White Paper}}, \href{https://doi.org/10.1088/0034-4885/70/11/R02}{\emph{Rept. Prog. Phys.} {\bfseries 70} (2007) 1757} [\href{https://arxiv.org/abs/hep-ph/0510213}{{\ttfamily hep-ph/0510213}}].

\bibitem{Drewes:2013gca}
M.~Drewes, \emph{{The Phenomenology of Right Handed Neutrinos}}, \href{https://doi.org/10.1142/S0218301313300191}{\emph{Int. J. Mod. Phys. E} {\bfseries 22} (2013) 1330019} [\href{https://arxiv.org/abs/1303.6912}{{\ttfamily 1303.6912}}].

\bibitem{Mohapatra:1986aw}
R.N.~Mohapatra, \emph{{Mechanism for Understanding Small Neutrino Mass in Superstring Theories}}, \href{https://doi.org/10.1103/PhysRevLett.56.561}{\emph{Phys. Rev. Lett.} {\bfseries 56} (1986) 561}.

\bibitem{Mohapatra:1986bd}
R.N.~Mohapatra and J.W.F.~Valle, \emph{{Neutrino Mass and Baryon Number Nonconservation in Superstring Models}}, \href{https://doi.org/10.1103/PhysRevD.34.1642}{\emph{Phys. Rev. D} {\bfseries 34} (1986) 1642}.

\bibitem{Bernabeu:1987gr}
J.~Bernabeu, A.~Santamaria, J.~Vidal, A.~Mendez and J.W.F.~Valle, \emph{{Lepton Flavor Nonconservation at High-Energies in a Superstring Inspired Standard Model}}, \href{https://doi.org/10.1016/0370-2693(87)91100-2}{\emph{Phys. Lett. B} {\bfseries 187} (1987) 303}.

\bibitem{Gavela:2009cd}
M.B.~Gavela, T.~Hambye, D.~Hernandez and P.~Hernandez, \emph{{Minimal Flavour Seesaw Models}}, \href{https://doi.org/10.1088/1126-6708/2009/09/038}{\emph{JHEP} {\bfseries 09} (2009) 038} [\href{https://arxiv.org/abs/0906.1461}{{\ttfamily 0906.1461}}].

\bibitem{Parida:2010wq}
M.K.~Parida and A.~Raychaudhuri, \emph{{Inverse see-saw, leptogenesis, observable proton decay and $\Delta^{\pm\pm}_{\rm R}$ in SUSY SO(10) with heavy W\_R}}, \href{https://doi.org/10.1103/PhysRevD.82.093017}{\emph{Phys. Rev. D} {\bfseries 82} (2010) 093017} [\href{https://arxiv.org/abs/1007.5085}{{\ttfamily 1007.5085}}].

\bibitem{Garayoa:2006xs}
J.~Garayoa, M.C.~Gonzalez-Garcia and N.~Rius, \emph{{Soft leptogenesis in the inverse seesaw model}}, \href{https://doi.org/10.1088/1126-6708/2007/02/021}{\emph{JHEP} {\bfseries 02} (2007) 021} [\href{https://arxiv.org/abs/hep-ph/0611311}{{\ttfamily hep-ph/0611311}}].

\bibitem{Abada:2014vea}
A.~Abada and M.~Lucente, \emph{{Looking for the minimal inverse seesaw realisation}}, \href{https://doi.org/10.1016/j.nuclphysb.2014.06.003}{\emph{Nucl. Phys. B} {\bfseries 885} (2014) 651} [\href{https://arxiv.org/abs/1401.1507}{{\ttfamily 1401.1507}}].

\bibitem{Law:2013gma}
S.S.C.~Law and K.L.~McDonald, \emph{{Generalized inverse seesaw mechanisms}}, \href{https://doi.org/10.1103/PhysRevD.87.113003}{\emph{Phys. Rev. D} {\bfseries 87} (2013) 113003} [\href{https://arxiv.org/abs/1303.4887}{{\ttfamily 1303.4887}}].

\bibitem{Nguyen:2020ehj}
T.P.~Nguyen, T.T.~Thuc, D.T.~Si, T.T.~Hong and L.T.~Hue, \emph{{Low energy phenomena of the lepton sector in an $A_4$ symmetry model with heavy inverse seesaw neutrinos}},  \href{https://arxiv.org/abs/2011.12181}{{\ttfamily 2011.12181}}.

\bibitem{Asaka:2005pn}
T.~Asaka and M.~Shaposhnikov, \emph{{The $\nu$MSM, dark matter and baryon asymmetry of the universe}}, \href{https://doi.org/10.1016/j.physletb.2005.06.020}{\emph{Phys. Lett. B} {\bfseries 620} (2005) 17} [\href{https://arxiv.org/abs/hep-ph/0505013}{{\ttfamily hep-ph/0505013}}].

\bibitem{Ma:2006fn}
E.~Ma, \emph{{Common origin of neutrino mass, dark matter, and baryogenesis}}, \href{https://doi.org/10.1142/S0217732306021141}{\emph{Mod. Phys. Lett. A} {\bfseries 21} (2006) 1777} [\href{https://arxiv.org/abs/hep-ph/0605180}{{\ttfamily hep-ph/0605180}}].

\bibitem{Falkowski:2011xh}
A.~Falkowski, J.T.~Ruderman and T.~Volansky, \emph{{Asymmetric Dark Matter from Leptogenesis}}, \href{https://doi.org/10.1007/JHEP05(2011)106}{\emph{JHEP} {\bfseries 05} (2011) 106} [\href{https://arxiv.org/abs/1101.4936}{{\ttfamily 1101.4936}}].

\bibitem{Falkowski:2017uya}
A.~Falkowski, E.~Kuflik, N.~Levi and T.~Volansky, \emph{{Light Dark Matter from Leptogenesis}}, \href{https://doi.org/10.1103/PhysRevD.99.015022}{\emph{Phys. Rev. D} {\bfseries 99} (2019) 015022} [\href{https://arxiv.org/abs/1712.07652}{{\ttfamily 1712.07652}}].

\bibitem{Hugle:2018qbw}
T.~Hugle, M.~Platscher and K.~Schmitz, \emph{{Low-Scale Leptogenesis in the Scotogenic Neutrino Mass Model}}, \href{https://doi.org/10.1103/PhysRevD.98.023020}{\emph{Phys. Rev. D} {\bfseries 98} (2018) 023020} [\href{https://arxiv.org/abs/1804.09660}{{\ttfamily 1804.09660}}].

\bibitem{Chianese:2019epo}
M.~Chianese, B.~Fu and S.F.~King, \emph{{Minimal Seesaw extension for Neutrino Mass and Mixing, Leptogenesis and Dark Matter: FIMPzillas through the Right-Handed Neutrino Portal}}, \href{https://doi.org/10.1088/1475-7516/2020/03/030}{\emph{JCAP} {\bfseries 03} (2020) 030} [\href{https://arxiv.org/abs/1910.12916}{{\ttfamily 1910.12916}}].

\bibitem{Liu:2020mxj}
A.~Liu, Z.-L.~Han, Y.~Jin and F.-X.~Yang, \emph{{Leptogenesis and dark matter from a low scale seesaw mechanism}}, \href{https://doi.org/10.1103/PhysRevD.101.095005}{\emph{Phys. Rev. D} {\bfseries 101} (2020) 095005} [\href{https://arxiv.org/abs/2001.04085}{{\ttfamily 2001.04085}}].

\bibitem{Fukugita:1986hr}
M.~Fukugita and T.~Yanagida, \emph{{Baryogenesis Without Grand Unification}}, \href{https://doi.org/10.1016/0370-2693(86)91126-3}{\emph{Phys. Lett. B} {\bfseries 174} (1986) 45}.

\bibitem{Covi:1996wh}
L.~Covi, E.~Roulet and F.~Vissani, \emph{{CP violating decays in leptogenesis scenarios}}, \href{https://doi.org/10.1016/0370-2693(96)00817-9}{\emph{Phys. Lett. B} {\bfseries 384} (1996) 169} [\href{https://arxiv.org/abs/hep-ph/9605319}{{\ttfamily hep-ph/9605319}}].

\bibitem{Roulet:1997xa}
E.~Roulet, L.~Covi and F.~Vissani, \emph{{On the CP asymmetries in Majorana neutrino decays}}, \href{https://doi.org/10.1016/S0370-2693(98)00135-X}{\emph{Phys. Lett. B} {\bfseries 424} (1998) 101} [\href{https://arxiv.org/abs/hep-ph/9712468}{{\ttfamily hep-ph/9712468}}].

\bibitem{Pilaftsis:1997jf}
A.~Pilaftsis, \emph{{CP violation and baryogenesis due to heavy Majorana neutrinos}}, \href{https://doi.org/10.1103/PhysRevD.56.5431}{\emph{Phys. Rev. D} {\bfseries 56} (1997) 5431} [\href{https://arxiv.org/abs/hep-ph/9707235}{{\ttfamily hep-ph/9707235}}].

\bibitem{Buchmuller:2005eh}
W.~Buchmuller, R.D.~Peccei and T.~Yanagida, \emph{{Leptogenesis as the origin of matter}}, \href{https://doi.org/10.1146/annurev.nucl.55.090704.151558}{\emph{Ann. Rev. Nucl. Part. Sci.} {\bfseries 55} (2005) 311} [\href{https://arxiv.org/abs/hep-ph/0502169}{{\ttfamily hep-ph/0502169}}].

\bibitem{Chun:2007vh}
E.J.~Chun and K.~Turzynski, \emph{{Quasi-degenerate neutrinos and leptogenesis from L(mu) - L(tau)}}, \href{https://doi.org/10.1103/PhysRevD.76.053008}{\emph{Phys. Rev. D} {\bfseries 76} (2007) 053008} [\href{https://arxiv.org/abs/hep-ph/0703070}{{\ttfamily hep-ph/0703070}}].

\bibitem{Kitabayashi:2007bs}
T.~Kitabayashi, \emph{{Remark on the minimal seesaw model and leptogenesis with tri/bi-maximal mixing}}, \href{https://doi.org/10.1103/PhysRevD.76.033002}{\emph{Phys. Rev. D} {\bfseries 76} (2007) 033002} [\href{https://arxiv.org/abs/hep-ph/0703303}{{\ttfamily hep-ph/0703303}}].

\bibitem{Prieto:2009zz}
C.~Martinez-Prieto, D.~Delepine and L.A.~Urena-Lopez, \emph{{Leptogenesis and Reheating in Complex Hybrid Inflation}}, \href{https://doi.org/10.1103/PhysRevD.81.036001}{\emph{Phys. Rev. D} {\bfseries 81} (2010) 036001} [\href{https://arxiv.org/abs/0908.2436}{{\ttfamily 0908.2436}}].

\bibitem{Suematsu:2011va}
D.~Suematsu, \emph{{Thermal Leptogenesis in a TeV Scale Model for Neutrino Masses}}, \href{https://doi.org/10.1140/epjc/s10052-012-1951-z}{\emph{Eur. Phys. J. C} {\bfseries 72} (2012) 1951} [\href{https://arxiv.org/abs/1103.0857}{{\ttfamily 1103.0857}}].

\bibitem{AristizabalSierra:2011ab}
D.~Aristizabal~Sierra, F.~Bazzocchi and I.~de~Medeiros~Varzielas, \emph{{Leptogenesis in flavor models with type I and II seesaws}}, \href{https://doi.org/10.1016/j.nuclphysb.2012.01.009}{\emph{Nucl. Phys. B} {\bfseries 858} (2012) 196} [\href{https://arxiv.org/abs/1112.1843}{{\ttfamily 1112.1843}}].

\bibitem{Hambye:2012fh}
T.~Hambye, \emph{{Leptogenesis: beyond the minimal type I seesaw scenario}}, \href{https://doi.org/10.1088/1367-2630/14/12/125014}{\emph{New J. Phys.} {\bfseries 14} (2012) 125014} [\href{https://arxiv.org/abs/1212.2888}{{\ttfamily 1212.2888}}].

\bibitem{Kashiwase:2013uy}
S.~Kashiwase and D.~Suematsu, \emph{{Leptogenesis and dark matter detection in a TeV scale neutrino mass model with inverted mass hierarchy}}, \href{https://doi.org/10.1140/epjc/s10052-013-2484-9}{\emph{Eur. Phys. J. C} {\bfseries 73} (2013) 2484} [\href{https://arxiv.org/abs/1301.2087}{{\ttfamily 1301.2087}}].

\bibitem{Borah:2013bza}
D.~Borah and M.K.~Das, \emph{{Neutrino Masses and Leptogenesis in Type I and Type II Seesaw Models}}, \href{https://doi.org/10.1103/PhysRevD.90.015006}{\emph{Phys. Rev. D} {\bfseries 90} (2014) 015006} [\href{https://arxiv.org/abs/1303.1758}{{\ttfamily 1303.1758}}].

\bibitem{Hamada:2015xva}
Y.~Hamada and K.~Kawana, \emph{{Reheating-era leptogenesis}}, \href{https://doi.org/10.1016/j.physletb.2016.10.067}{\emph{Phys. Lett. B} {\bfseries 763} (2016) 388} [\href{https://arxiv.org/abs/1510.05186}{{\ttfamily 1510.05186}}].

\bibitem{Zhao:2020bzx}
Z.-h.~Zhao, \emph{{Renormalization group evolution induced leptogenesis in the minimal seesaw model with the trimaximal mixing and mu-tau reflection symmetry}}, \href{https://doi.org/10.1007/JHEP11(2021)170}{\emph{JHEP} {\bfseries 11} (2021) 170} [\href{https://arxiv.org/abs/2003.00654}{{\ttfamily 2003.00654}}].

\bibitem{Davidson:2008bu}
S.~Davidson, E.~Nardi and Y.~Nir, \emph{{Leptogenesis}}, \href{https://doi.org/10.1016/j.physrep.2008.06.002}{\emph{Phys. Rept.} {\bfseries 466} (2008) 105} [\href{https://arxiv.org/abs/0802.2962}{{\ttfamily 0802.2962}}].

\bibitem{Blanchet:2012bk}
S.~Blanchet and P.~Di~Bari, \emph{{The minimal scenario of leptogenesis}}, \href{https://doi.org/10.1088/1367-2630/14/12/125012}{\emph{New J. Phys.} {\bfseries 14} (2012) 125012} [\href{https://arxiv.org/abs/1211.0512}{{\ttfamily 1211.0512}}].

\bibitem{Manton:1983nd}
N.S.~Manton, \emph{{Topology in the Weinberg-Salam Theory}}, \href{https://doi.org/10.1103/PhysRevD.28.2019}{\emph{Phys. Rev. D} {\bfseries 28} (1983) 2019}.

\bibitem{Klinkhamer:1984di}
F.R.~Klinkhamer and N.S.~Manton, \emph{{A Saddle Point Solution in the Weinberg-Salam Theory}}, \href{https://doi.org/10.1103/PhysRevD.30.2212}{\emph{Phys. Rev. D} {\bfseries 30} (1984) 2212}.

\bibitem{Kuzmin:1985mm}
V.A.~Kuzmin, V.A.~Rubakov and M.E.~Shaposhnikov, \emph{{On the Anomalous Electroweak Baryon Number Nonconservation in the Early Universe}}, \href{https://doi.org/10.1016/0370-2693(85)91028-7}{\emph{Phys. Lett. B} {\bfseries 155} (1985) 36}.

\bibitem{Khlebnikov:1988sr}
S.Y.~Khlebnikov and M.E.~Shaposhnikov, \emph{{The Statistical Theory of Anomalous Fermion Number Nonconservation}}, \href{https://doi.org/10.1016/0550-3213(88)90133-2}{\emph{Nucl. Phys. B} {\bfseries 308} (1988) 885}.

\bibitem{Harvey:1990qw}
J.A.~Harvey and M.S.~Turner, \emph{{Cosmological Baryon and Lepton Number in the Presence of Electroweak Fermion Number Violation}}, \href{https://doi.org/10.1103/PhysRevD.42.3344}{\emph{Phys. Rev. D} {\bfseries 42} (1990) 3344}.

\bibitem{Davidson:2002qv}
S.~Davidson and A.~Ibarra, \emph{{A Lower bound on the right-handed neutrino mass from leptogenesis}}, \href{https://doi.org/10.1016/S0370-2693(02)01735-5}{\emph{Phys. Lett. B} {\bfseries 535} (2002) 25} [\href{https://arxiv.org/abs/hep-ph/0202239}{{\ttfamily hep-ph/0202239}}].

\bibitem{Flanz:1996fb}
M.~Flanz, E.A.~Paschos, U.~Sarkar and J.~Weiss, \emph{{Baryogenesis through mixing of heavy Majorana neutrinos}}, \href{https://doi.org/10.1016/S0370-2693(96)01337-8}{\emph{Phys. Lett. B} {\bfseries 389} (1996) 693} [\href{https://arxiv.org/abs/hep-ph/9607310}{{\ttfamily hep-ph/9607310}}].

\bibitem{Pilaftsis:1997dr}
A.~Pilaftsis, \emph{{Resonant CP violation induced by particle mixing in transition amplitudes}}, \href{https://doi.org/10.1016/S0550-3213(97)00469-0}{\emph{Nucl. Phys. B} {\bfseries 504} (1997) 61} [\href{https://arxiv.org/abs/hep-ph/9702393}{{\ttfamily hep-ph/9702393}}].

\bibitem{Pilaftsis:2003gt}
A.~Pilaftsis and T.E.J.~Underwood, \emph{{Resonant leptogenesis}}, \href{https://doi.org/10.1016/j.nuclphysb.2004.05.029}{\emph{Nucl. Phys. B} {\bfseries 692} (2004) 303} [\href{https://arxiv.org/abs/hep-ph/0309342}{{\ttfamily hep-ph/0309342}}].

\bibitem{Dev:2017wwc}
B.~Dev, M.~Garny, J.~Klaric, P.~Millington and D.~Teresi, \emph{{Resonant enhancement in leptogenesis}}, \href{https://doi.org/10.1142/S0217751X18420034}{\emph{Int. J. Mod. Phys. A} {\bfseries 33} (2018) 1842003} [\href{https://arxiv.org/abs/1711.02863}{{\ttfamily 1711.02863}}].

\bibitem{Chakraborty:2021azg}
I.~Chakraborty, H.~Roy and T.~Srivastava, \emph{{Resonant leptogenesis in (2,2) inverse see-saw realisation}}, \href{https://doi.org/10.1016/j.nuclphysb.2022.115780}{\emph{Nucl. Phys. B} {\bfseries 979} (2022) 115780} [\href{https://arxiv.org/abs/2106.08232}{{\ttfamily 2106.08232}}].

\bibitem{Chakraborty:2022pcc}
I.~Chakraborty, H.~Roy and T.~Srivastava, \emph{{Searches for heavy neutrinos at multi-TeV muon collider: a resonant leptogenesis perspective}}, \href{https://doi.org/10.1140/epjc/s10052-023-11406-0}{\emph{Eur. Phys. J. C} {\bfseries 83} (2023) 280} [\href{https://arxiv.org/abs/2206.07037}{{\ttfamily 2206.07037}}].

\bibitem{Deppisch:2015qwa}
F.F.~Deppisch, P.S.~Bhupal~Dev and A.~Pilaftsis, \emph{{Neutrinos and Collider Physics}}, \href{https://doi.org/10.1088/1367-2630/17/7/075019}{\emph{New J. Phys.} {\bfseries 17} (2015) 075019} [\href{https://arxiv.org/abs/1502.06541}{{\ttfamily 1502.06541}}].

\bibitem{Cai:2017mow}
Y.~Cai, T.~Han, T.~Li and R.~Ruiz, \emph{{Lepton Number Violation: Seesaw Models and Their Collider Tests}}, \href{https://doi.org/10.3389/fphy.2018.00040}{\emph{Front. in Phys.} {\bfseries 6} (2018) 40} [\href{https://arxiv.org/abs/1711.02180}{{\ttfamily 1711.02180}}].

\bibitem{deGouvea:2013zba}
A.~de~Gouvea and P.~Vogel, \emph{{Lepton Flavor and Number Conservation, and Physics Beyond the Standard Model}}, \href{https://doi.org/10.1016/j.ppnp.2013.03.006}{\emph{Prog. Part. Nucl. Phys.} {\bfseries 71} (2013) 75} [\href{https://arxiv.org/abs/1303.4097}{{\ttfamily 1303.4097}}].

\bibitem{Alekhin:2015byh}
S.~Alekhin et~al., \emph{{A facility to Search for Hidden Particles at the CERN SPS: the SHiP physics case}}, \href{https://doi.org/10.1088/0034-4885/79/12/124201}{\emph{Rept. Prog. Phys.} {\bfseries 79} (2016) 124201} [\href{https://arxiv.org/abs/1504.04855}{{\ttfamily 1504.04855}}].

\bibitem{Bertone:2004pz}
G.~Bertone, D.~Hooper and J.~Silk, \emph{{Particle dark matter: Evidence, candidates and constraints}}, \href{https://doi.org/10.1016/j.physrep.2004.08.031}{\emph{Phys. Rept.} {\bfseries 405} (2005) 279} [\href{https://arxiv.org/abs/hep-ph/0404175}{{\ttfamily hep-ph/0404175}}].

\bibitem{Cirelli:2024ssz}
M.~Cirelli, A.~Strumia and J.~Zupan, \emph{{Dark Matter}},  \href{https://arxiv.org/abs/2406.01705}{{\ttfamily 2406.01705}}.

\bibitem{Planck:2018vyg}
{\scshape Planck} collaboration, \emph{{Planck 2018 results. VI. Cosmological parameters}}, \href{https://doi.org/10.1051/0004-6361/201833910}{\emph{Astron. Astrophys.} {\bfseries 641} (2020) A6} [\href{https://arxiv.org/abs/1807.06209}{{\ttfamily 1807.06209}}].

\bibitem{Liu:2024esf}
A.~Liu, F.-L.~Shao, Z.-L.~Han, Y.~Jin and H.~Li, \emph{{Common origin of dark matter and leptogenesis in U(1)$_{B-L}$}}, \href{https://doi.org/10.1007/JHEP10(2024)019}{\emph{JHEP} {\bfseries 10} (2024) 019} [\href{https://arxiv.org/abs/2407.19730}{{\ttfamily 2407.19730}}].

\bibitem{Marshak:1979fm}
R.E.~Marshak and R.N.~Mohapatra, \emph{{Quark - Lepton Symmetry and B-L as the U(1) Generator of the Electroweak Symmetry Group}}, \href{https://doi.org/10.1016/0370-2693(80)90436-0}{\emph{Phys. Lett. B} {\bfseries 91} (1980) 222}.

\bibitem{Mohapatra:1980qe}
R.N.~Mohapatra and R.E.~Marshak, \emph{{Local B-L Symmetry of Electroweak Interactions, Majorana Neutrinos and Neutron Oscillations}}, \href{https://doi.org/10.1103/PhysRevLett.44.1316}{\emph{Phys. Rev. Lett.} {\bfseries 44} (1980) 1316}.

\bibitem{Okada:2010wd}
N.~Okada and O.~Seto, \emph{{Higgs portal dark matter in the minimal gauged $U(1)_{B-L}$ model}}, \href{https://doi.org/10.1103/PhysRevD.82.023507}{\emph{Phys. Rev. D} {\bfseries 82} (2010) 023507} [\href{https://arxiv.org/abs/1002.2525}{{\ttfamily 1002.2525}}].

\bibitem{Escudero:2016tzx}
M.~Escudero, N.~Rius and V.~Sanz, \emph{{Sterile neutrino portal to Dark Matter I: The $U(1)_{B-L}$ case}}, \href{https://doi.org/10.1007/JHEP02(2017)045}{\emph{JHEP} {\bfseries 02} (2017) 045} [\href{https://arxiv.org/abs/1606.01258}{{\ttfamily 1606.01258}}].

\bibitem{Okada:2018ktp}
S.~Okada, \emph{{$Z'$ Portal Dark Matter in the Minimal $B-L$ Model}}, \href{https://doi.org/10.1155/2018/5340935}{\emph{Adv. High Energy Phys.} {\bfseries 2018} (2018) 5340935} [\href{https://arxiv.org/abs/1803.06793}{{\ttfamily 1803.06793}}].

\bibitem{Iso:2010mv}
S.~Iso, N.~Okada and Y.~Orikasa, \emph{{Resonant Leptogenesis in the Minimal B-L Extended Standard Model at TeV}}, \href{https://doi.org/10.1103/PhysRevD.83.093011}{\emph{Phys. Rev. D} {\bfseries 83} (2011) 093011} [\href{https://arxiv.org/abs/1011.4769}{{\ttfamily 1011.4769}}].

\bibitem{Heeck:2016oda}
J.~Heeck and D.~Teresi, \emph{{Leptogenesis and neutral gauge bosons}}, \href{https://doi.org/10.1103/PhysRevD.94.095024}{\emph{Phys. Rev. D} {\bfseries 94} (2016) 095024} [\href{https://arxiv.org/abs/1609.03594}{{\ttfamily 1609.03594}}].

\bibitem{Dev:2017xry}
P.S.B.~Dev, R.N.~Mohapatra and Y.~Zhang, \emph{{Leptogenesis constraints on $B - L$ breaking Higgs boson in TeV scale seesaw models}}, \href{https://doi.org/10.1007/JHEP03(2018)122}{\emph{JHEP} {\bfseries 03} (2018) 122} [\href{https://arxiv.org/abs/1711.07634}{{\ttfamily 1711.07634}}].

\bibitem{Han:2021udl}
T.~Han, S.~Li, S.~Su, W.~Su and Y.~Wu, \emph{{Heavy Higgs bosons in 2HDM at a muon collider}}, \href{https://doi.org/10.1103/PhysRevD.104.055029}{\emph{Phys. Rev. D} {\bfseries 104} (2021) 055029} [\href{https://arxiv.org/abs/2102.08386}{{\ttfamily 2102.08386}}].

\bibitem{Esteban:2024eli}
I.~Esteban, M.C.~Gonzalez-Garcia, M.~Maltoni, I.~Martinez-Soler, J.P.~Pinheiro and T.~Schwetz, \emph{{NuFit-6.0: updated global analysis of three-flavor neutrino oscillations}}, \href{https://doi.org/10.1007/JHEP12(2024)216}{\emph{JHEP} {\bfseries 12} (2024) 216} [\href{https://arxiv.org/abs/2410.05380}{{\ttfamily 2410.05380}}].

\bibitem{TheMEG:2016wtm}
{\scshape MEG} collaboration, \emph{{Search for the lepton flavour violating decay $\mu ^+ \rightarrow \mathrm {e}^+ \gamma $ with the full dataset of the MEG experiment}}, \href{https://doi.org/10.1140/epjc/s10052-016-4271-x}{\emph{Eur. Phys. J. C} {\bfseries 76} (2016) 434} [\href{https://arxiv.org/abs/1605.05081}{{\ttfamily 1605.05081}}].

\bibitem{Aubert:2009ag}
{\scshape BaBar} collaboration, \emph{{Searches for Lepton Flavor Violation in the Decays $\tau^\pm \to e^\pm \gamma$ and $\tau^\pm \to \mu^\pm \gamma$}}, \href{https://doi.org/10.1103/PhysRevLett.104.021802}{\emph{Phys. Rev. Lett.} {\bfseries 104} (2010) 021802} [\href{https://arxiv.org/abs/0908.2381}{{\ttfamily 0908.2381}}].

\bibitem{Plumacher:1996kc}
M.~Plumacher, \emph{{Baryogenesis and lepton number violation}}, \href{https://doi.org/10.1007/s002880050418}{\emph{Z. Phys. C} {\bfseries 74} (1997) 549} [\href{https://arxiv.org/abs/hep-ph/9604229}{{\ttfamily hep-ph/9604229}}].

\bibitem{Fong:2021tqj}
C.S.~Fong, M.H.~Rahat and S.~Saad, \emph{{Low-scale resonant leptogenesis in SU(5) GUT with T13 family symmetry}}, \href{https://doi.org/10.1103/PhysRevD.104.095028}{\emph{Phys. Rev. D} {\bfseries 104} (2021) 095028} [\href{https://arxiv.org/abs/2103.14691}{{\ttfamily 2103.14691}}].

\bibitem{Buchmuller:2004nz}
W.~Buchmuller, P.~Di~Bari and M.~Plumacher, \emph{{Leptogenesis for pedestrians}}, \href{https://doi.org/10.1016/j.aop.2004.02.003}{\emph{Annals Phys.} {\bfseries 315} (2005) 305} [\href{https://arxiv.org/abs/hep-ph/0401240}{{\ttfamily hep-ph/0401240}}].

\bibitem{Buchmuller:2002rq}
W.~Buchmuller, P.~Di~Bari and M.~Plumacher, \emph{{Cosmic microwave background, matter - antimatter asymmetry and neutrino masses}}, \href{https://doi.org/10.1016/S0550-3213(02)00737-X}{\emph{Nucl. Phys. B} {\bfseries 643} (2002) 367} [\href{https://arxiv.org/abs/hep-ph/0205349}{{\ttfamily hep-ph/0205349}}].

\bibitem{Giudice:2003jh}
G.~Giudice, A.~Notari, M.~Raidal, A.~Riotto and A.~Strumia, \emph{{Towards a complete theory of thermal leptogenesis in the SM and MSSM}}, \href{https://doi.org/10.1016/j.nuclphysb.2004.02.019}{\emph{Nucl. Phys. B} {\bfseries 685} (2004) 89} [\href{https://arxiv.org/abs/hep-ph/0310123}{{\ttfamily hep-ph/0310123}}].

\bibitem{Chakraborty:2019zas}
I.~Chakraborty and H.~Roy, \emph{{Type-I thermal leptogenesis in $Z_3$-symmetric three Higgs doublet model}}, \href{https://doi.org/10.1140/epjc/s10052-020-8377-9}{\emph{Eur. Phys. J. C} {\bfseries 80} (2020) 1038} [\href{https://arxiv.org/abs/1909.07790}{{\ttfamily 1909.07790}}].

\bibitem{Rahat:2020mio}
M.H.~Rahat, \emph{{Leptogenesis from the Asymmetric Texture}}, \href{https://doi.org/10.1103/PhysRevD.103.035011}{\emph{Phys. Rev. D} {\bfseries 103} (2021) 035011} [\href{https://arxiv.org/abs/2008.04204}{{\ttfamily 2008.04204}}].

\bibitem{Garny:2011hg}
M.~Garny, A.~Kartavtsev and A.~Hohenegger, \emph{{Leptogenesis from first principles in the resonant regime}}, \href{https://doi.org/10.1016/j.aop.2012.10.007}{\emph{Annals Phys.} {\bfseries 328} (2013) 26} [\href{https://arxiv.org/abs/1112.6428}{{\ttfamily 1112.6428}}].

\bibitem{Iso:2013lba}
S.~Iso, K.~Shimada and M.~Yamanaka, \emph{{Kadanoff-Baym approach to the thermal resonant leptogenesis}}, \href{https://doi.org/10.1007/JHEP04(2014)062}{\emph{JHEP} {\bfseries 04} (2014) 062} [\href{https://arxiv.org/abs/1312.7680}{{\ttfamily 1312.7680}}].

\bibitem{Iso:2014afa}
S.~Iso and K.~Shimada, \emph{{Coherent Flavour Oscillation and CP Violating Parameter in Thermal Resonant Leptogenesis}}, \href{https://doi.org/10.1007/JHEP08(2014)043}{\emph{JHEP} {\bfseries 08} (2014) 043} [\href{https://arxiv.org/abs/1404.4816}{{\ttfamily 1404.4816}}].

\bibitem{Hambye:2001eu}
T.~Hambye, \emph{{Leptogenesis at the TeV scale}}, \href{https://doi.org/10.1016/S0550-3213(02)00293-6}{\emph{Nucl. Phys. B} {\bfseries 633} (2002) 171} [\href{https://arxiv.org/abs/hep-ph/0111089}{{\ttfamily hep-ph/0111089}}].

\bibitem{Hambye:2004jf}
T.~Hambye, J.~March-Russell and S.M.~West, \emph{{TeV scale resonant leptogenesis from supersymmetry breaking}}, \href{https://doi.org/10.1088/1126-6708/2004/07/070}{\emph{JHEP} {\bfseries 07} (2004) 070} [\href{https://arxiv.org/abs/hep-ph/0403183}{{\ttfamily hep-ph/0403183}}].

\bibitem{Pilaftsis:1998pd}
A.~Pilaftsis, \emph{{Heavy Majorana neutrinos and baryogenesis}}, \href{https://doi.org/10.1142/S0217751X99000932}{\emph{Int. J. Mod. Phys. A} {\bfseries 14} (1999) 1811} [\href{https://arxiv.org/abs/hep-ph/9812256}{{\ttfamily hep-ph/9812256}}].

\bibitem{Branco:2006hz}
G.C.~Branco, A.J.~Buras, S.~Jager, S.~Uhlig and A.~Weiler, \emph{{Another look at minimal lepton flavour violation, $l_i \to l_{j\gamma}$, leptogenesis, and the ratio $M_\nu / \Lambda_{LFV}$}}, \href{https://doi.org/10.1088/1126-6708/2007/09/004}{\emph{JHEP} {\bfseries 09} (2007) 004} [\href{https://arxiv.org/abs/hep-ph/0609067}{{\ttfamily hep-ph/0609067}}].

\bibitem{Brdar:2019iem}
V.~Brdar, A.J.~Helmboldt, S.~Iwamoto and K.~Schmitz, \emph{{Type-I Seesaw as the Common Origin of Neutrino Mass, Baryon Asymmetry, and the Electroweak Scale}}, \href{https://doi.org/10.1103/PhysRevD.100.075029}{\emph{Phys. Rev. D} {\bfseries 100} (2019) 075029} [\href{https://arxiv.org/abs/1905.12634}{{\ttfamily 1905.12634}}].

\bibitem{Burnier:2005hp}
Y.~Burnier, M.~Laine and M.~Shaposhnikov, \emph{{Baryon and lepton number violation rates across the electroweak crossover}}, \href{https://doi.org/10.1088/1475-7516/2006/02/007}{\emph{JCAP} {\bfseries 02} (2006) 007} [\href{https://arxiv.org/abs/hep-ph/0511246}{{\ttfamily hep-ph/0511246}}].

\bibitem{Belanger:2018mqt}
G.~B{\'e}langer, F.~Boudjema, A.~Goudelis, A.~Pukhov and B.~Zaldivar, \emph{{micrOMEGAs5.0 : Freeze-in}}, \href{https://doi.org/10.1016/j.cpc.2018.04.027}{\emph{Comput. Phys. Commun.} {\bfseries 231} (2018) 173} [\href{https://arxiv.org/abs/1801.03509}{{\ttfamily 1801.03509}}].

\bibitem{LZ:2024zvo}
{\scshape LZ} collaboration, \emph{{Dark Matter Search Results from 4.2{\,}{\,}Tonne-Years of Exposure of the LUX-ZEPLIN (LZ) Experiment}}, \href{https://doi.org/10.1103/4dyc-z8zf}{\emph{Phys. Rev. Lett.} {\bfseries 135} (2025) 011802} [\href{https://arxiv.org/abs/2410.17036}{{\ttfamily 2410.17036}}].

\bibitem{Borah:2021mri}
D.~Borah, A.~Dasgupta and D.~Mahanta, \emph{{TeV scale resonant leptogenesis with $L_{\mu}-L_{\tau}$ gauge symmetry in light of the muon $g-2$}}, \href{https://doi.org/10.1103/PhysRevD.104.075006}{\emph{Phys. Rev. D} {\bfseries 104} (2021) 075006} [\href{https://arxiv.org/abs/2106.14410}{{\ttfamily 2106.14410}}].

\bibitem{ATLAS:2019erb}
{\scshape ATLAS} collaboration, \emph{{Search for high-mass dilepton resonances using 139 fb$^{-1}$ of $pp$ collision data collected at $\sqrt{s}=$13 TeV with the ATLAS detector}}, \href{https://doi.org/10.1016/j.physletb.2019.07.016}{\emph{Phys. Lett. B} {\bfseries 796} (2019) 68} [\href{https://arxiv.org/abs/1903.06248}{{\ttfamily 1903.06248}}].

\bibitem{CMS:2021ctt}
{\scshape CMS} collaboration, \emph{{Search for resonant and nonresonant new phenomena in high-mass dilepton final states at $ \sqrt{s} $ = 13 TeV}}, \href{https://doi.org/10.1007/JHEP07(2021)208}{\emph{JHEP} {\bfseries 07} (2021) 208} [\href{https://arxiv.org/abs/2103.02708}{{\ttfamily 2103.02708}}].

\bibitem{Homiller:2022iax}
S.~Homiller, Q.~Lu and M.~Reece, \emph{{Complementary signals of lepton flavor violation at a high-energy muon collider}}, \href{https://doi.org/10.1007/JHEP07(2022)036}{\emph{JHEP} {\bfseries 07} (2022) 036} [\href{https://arxiv.org/abs/2203.08825}{{\ttfamily 2203.08825}}].

\bibitem{Ally:2022rgk}
D.~Ally, L.~Carpenter, T.~Holmes, L.~Lee and P.~Wagenknecht, \emph{{Strategies for Beam-Induced Background Reduction at Muon Colliders}},  in \emph{{Snowmass 2021}}, 3, 2022 [\href{https://arxiv.org/abs/2203.06773}{{\ttfamily 2203.06773}}].

\bibitem{Antonelli:2015nla}
M.~Antonelli, M.~Boscolo, R.~Di~Nardo and P.~Raimondi, \emph{{Novel proposal for a low emittance muon beam using positron beam on target}}, \href{https://doi.org/10.1016/j.nima.2015.10.097}{\emph{Nucl. Instrum. Meth. A} {\bfseries 807} (2016) 101} [\href{https://arxiv.org/abs/1509.04454}{{\ttfamily 1509.04454}}].

\bibitem{Alwall:2014hca}
J.~Alwall, R.~Frederix, S.~Frixione, V.~Hirschi, F.~Maltoni, O.~Mattelaer et~al., \emph{{The automated computation of tree-level and next-to-leading order differential cross sections, and their matching to parton shower simulations}}, \href{https://doi.org/10.1007/JHEP07(2014)079}{\emph{JHEP} {\bfseries 07} (2014) 079} [\href{https://arxiv.org/abs/1405.0301}{{\ttfamily 1405.0301}}].

\bibitem{Sjostrand:2014zea}
T.~Sj{\"o}strand, S.~Ask, J.R.~Christiansen, R.~Corke, N.~Desai, P.~Ilten et~al., \emph{{An introduction to PYTHIA 8.2}}, \href{https://doi.org/10.1016/j.cpc.2015.01.024}{\emph{Comput. Phys. Commun.} {\bfseries 191} (2015) 159} [\href{https://arxiv.org/abs/1410.3012}{{\ttfamily 1410.3012}}].

\bibitem{Bierlich:2022pfr}
C.~Bierlich et~al., \emph{{A comprehensive guide to the physics and usage of PYTHIA 8.3}}, \href{https://doi.org/10.21468/SciPostPhysCodeb.8}{\emph{SciPost Phys. Codeb.} {\bfseries 2022} (2022) 8} [\href{https://arxiv.org/abs/2203.11601}{{\ttfamily 2203.11601}}].

\bibitem{deFavereau:2013fsa}
{\scshape DELPHES 3} collaboration, \emph{{DELPHES 3, A modular framework for fast simulation of a generic collider experiment}}, \href{https://doi.org/10.1007/JHEP02(2014)057}{\emph{JHEP} {\bfseries 02} (2014) 057} [\href{https://arxiv.org/abs/1307.6346}{{\ttfamily 1307.6346}}].

\bibitem{Boronat:2016tgd}
M.~Boronat, J.~Fuster, I.~Garcia, P.~Roloff, R.~Simoniello and M.~Vos, \emph{{Jet reconstruction at high-energy electron{\textendash}positron colliders}}, \href{https://doi.org/10.1140/epjc/s10052-018-5594-6}{\emph{Eur. Phys. J. C} {\bfseries 78} (2018) 144} [\href{https://arxiv.org/abs/1607.05039}{{\ttfamily 1607.05039}}].

\bibitem{Boronat:2014hva}
M.~Boronat, J.~Fuster, I.~Garcia, E.~Ros and M.~Vos, \emph{{A robust jet reconstruction algorithm for high-energy lepton colliders}}, \href{https://doi.org/10.1016/j.physletb.2015.08.055}{\emph{Phys. Lett. B} {\bfseries 750} (2015) 95} [\href{https://arxiv.org/abs/1404.4294}{{\ttfamily 1404.4294}}].

\end{thebibliography}\endgroup
\end{document}